\documentclass[conference]{IEEEtran}

\usepackage{cite}
\usepackage[utf8]{inputenc}
\usepackage{csquotes}
\usepackage{tabularx}
\usepackage{booktabs}
\usepackage{multirow}
\usepackage{threeparttable}
\usepackage{colortbl}
\usepackage{algorithm}
\usepackage{algpseudocode}
\usepackage{pgf}
\usepackage{tikz}
\usepackage{lscape}
\usepackage{pgfplots}
\usepackage{etoolbox}
\usepackage[simplified]{pgf-umlcd}
\usepackage{pgfplotstable}
\usepgfplotslibrary{fillbetween}
\usepackage{hyperref}
\usepackage[normalem]{ulem}
\usepackage{amsmath, amssymb, amsfonts}
\usepackage{graphicx}
\usepackage{textcomp}
\usepackage{xcolor}
\usepackage{subcaption}
\usepackage{import} % for inkscape pdf_tex file imports
\usepackage{listings}
\usepackage{xfrac} % for pretty fractions
\usepackage{amsthm} % for unnumbered theorem-like stuff
\usepackage{fancyvrb}
\usepackage{stmaryrd}
\usetikzlibrary{decorations.text}
\usepackage[dvipsnames, table]{xcolor}
\usepackage[most]{tcolorbox} % for boxes around equations
\usepackage{diagbox}
\usepackage{soul}
\usepackage{csquotes}
\usepackage{graphicx}
\usepackage{url}
\usepackage[export]{adjustbox}
\usepackage{ifthen}

\usetikzlibrary{math} 
\usetikzlibrary{patterns}
\usetikzlibrary{decorations.pathmorphing}
\usetikzlibrary{arrows.meta,automata,shadows,shapes,shapes.geometric,circuits.logic.US,shapes.gates.logic.US,calc}

\tikzset{
    circuit logic US,
	>=Stealth, % can't stand the default arrows
	multrectangle/.style={draw, fill=black!5},
	generator/.style={draw,align=center,fill=blue!5,rounded corners, minimum width=15ex, minimum height=8ex},
    hwblock/.style={draw, rectangle, rounded corners=.3, thick, fill=black!5, drop shadow={shadow xshift=.5ex,shadow yshift=-.5ex}, font=\sffamily, minimum height=5ex,align=center},
	hwregblock/.style={hwblock, fill=blue!10},
	every circuit symbol/.style={hwblock},
	filteradd/.style={hwblock, circle, minimum height=1ex},
	filtermult/.style={hwblock, minimum height=7ex, regular polygon, regular polygon sides=3, shape border rotate=180, inner sep= .2ex},
	hwbus/.style={very thick,>={Stealth[length=6pt]}},
	hwbuslarge/.style={line width=2pt,>={Stealth[length=7pt]}},
	hwbuswhitebackground/.style={line width=4pt,color=white},
	hwwire/.style={thin, >={Stealth[length=4pt]}  },
	hwword/.style={draw, rectangle, minimum height=3ex},
	bitwidth/.style={font=\scriptsize, red}, % Turn back to blue in the final version?
	abitwidth/.style={bitwidth,above},
	bbitwidth/.style={bitwidth,below},
	lbitwidth/.style={bitwidth,left},
	rbitwidth/.style={bitwidth,right},
	flopocoClass/.style={font=\small\tt,draw,align=center,rectangle,fill=blue!5,minimum height=3ex},
	flopocoInfo/.style={font=\small,align=center,rectangle, fill=yellow!50,fill opacity=0.5, text opacity=1,rounded corners=10}
}
\newboolean{showbitwidths} %  we can remove bitwidth them globally, don't remember the reason
\setboolean{showbitwidths}{true}
\newcommand{\bitwidth}[3]{ % coordinate, text, [a|b|l|r]
	\ifthenelse{\boolean{showbitwidths}}{
		\draw[bitwidth,ultra thin] (#1) ++(0.5ex, 0.5ex)  -- ++(-1ex, -1ex) ++(0.5ex, 0.5ex)  node[#3bitwidth] {#2};
	}{}
} %

\pgfmathsetmacro{\nodeinputwirelength}{5}
\pgfmathsetmacro{\nodeinputshiftpos}{.4}
\pgfmathsetmacro{\nodeoutputwirelength}{1}
\pgfmathsetmacro{\nodeoutputshiftwirelength}{5}
\pgfmathsetmacro{\nodeoutputshiftpos}{.7}
\pgfmathsetmacro{\nodestage}{2.5} % stage vertical spacing (from one input to center of stage above, see usage example below)
\pgfmathsetmacro{\prevstagedeltayfrominput}{2} %  see usage example below)
\pgfmathsetmacro{\prevstagedeltayfromnode}{8} %  see usage example below)
\newcommand{\shift}[2]{%
    \ifnum#2<0
        \edef\absval{\the\numexpr-#2\relax}%
        \draw[thick, ->, >=latex, blue] (#1) ++(-1,0) -- ++(3,0) node[pos=.7,above, fill=white, inner sep=1pt] {\small \absval};
    \else
        \edef\absval{\the\numexpr#2\relax}%
        \draw[thick, ->, >=latex, blue] (#1) ++(1,0) -- ++(-3,0) node[pos=.7,above, fill=white, inner sep=1pt] {\small \absval};
    \fi
}
\newcommand{\shiftRight}[2]{%
    \ifnum#2<0
        \edef\absval{\the\numexpr-#2\relax}%
          \draw[thick, ->, >=latex, blue] (#1) ++(-1,0) -- ++(3,0) node[pos=0.5,above right, fill=white, inner sep=1pt] {\small \absval};
    \else
        \edef\absval{\the\numexpr#2\relax}%
        \draw[thick, ->, >=latex, blue] (#1) ++(1,0) -- ++(-3,0) node[pos=0.5,above right, fill=white, inner sep=1pt] {\small \absval};
    \fi
}
\newcommand{\shiftLeft}[2]{%
  \ifnum#2<0
    \edef\absval{\the\numexpr-#2\relax}%
    \draw[thick, ->, >=latex, blue] (#1) ++(-1,0) -- ++(3,0) node[pos=0.3,above left, fill=white, inner sep=1pt] {\small \absval};
  \else
    \edef\absval{\the\numexpr#2\relax}%
    \draw[thick, ->, >=latex, blue] (#1) ++(1,0) -- ++(-3,0) node[pos=0.3,above left, fill=white, inner sep=1pt] {\small \absval};
  \fi
}

\newcommand{\nodedrawio}[6]{ % position, name, leftshift, rightshift, outputshift, large shift name
  \ifthenelse{\equal{#3}{0} \AND \equal{#4}{0}} {
    \draw[hwbus, <-](#2i0) -- ++ (0,2) coordinate (#2i0); % renaming the coordinate;
    \draw[hwbus, <-](#2i1) -- ++ (0,2) coordinate (#2i1); % renaming the coordinate
  }{
    \draw[hwbus, <-](#2i0) -- ++ (0,\nodeinputwirelength)
      coordinate[pos=\nodeinputshiftpos](#2i0shift) coordinate (#2i0); % renaming the coordinate;
      \draw[hwbus, <-](#2i1) -- ++ (0,\nodeinputwirelength)
    coordinate[pos=\nodeinputshiftpos](#2i1shift) coordinate (#2i1); % renaming the coordinate
  }
    
  \ifthenelse{\equal{#6}{0}} {
    \ifthenelse{\equal{#3}{} \OR \equal{#3}{0}}{}{
      \shift{#2i0shift}{#3};
      }
    \ifthenelse{\equal{#4}{} \OR \equal{#4}{0}}{}{
      \shift{#2i1shift}{#4};
      }
  }{
    \ifthenelse{\equal{#3}{} \OR \equal{#3}{0}}{}{
      \shiftLeft{#2i0shift}{#3};
      }
    \ifthenelse{\equal{#4}{} \OR \equal{#4}{0}}{}{
      \shiftRight{#2i1shift}{#4};
      }
  }
  \ifthenelse{\equal{#5}{0}} {
    \draw[hwbus] (#2o) -- ++(0,-\nodeoutputwirelength) coordinate (#2o); % renaming the coordinate;
  }{
    \draw[hwbus] (#2o) -- ++ (0,-\nodeoutputshiftwirelength)
    coordinate[pos=\nodeoutputshiftpos](#2oshift) coordinate (#2o); % renaming the coordinate
    \ifthenelse{\equal{#6}{0}} {
      \shift{#2oshift}{#5};
    }{
      \shiftLeft{#2oshift}{#5};
    }
  }
}
\iftrue % Square adders
\newcommand{\customOperator}[3]{ % position, name, operator
  \draw(#1) node(#2)[hwblock, minimum width=6ex, minimum height=3ex]{$\pm$};
  \ifthenelse{\boolean{tikzdebug}}{  \draw(#2.south west) ++(0,.4)node[blue,fill=yellow]{\tiny #2}; }{}
  \draw(#2.south) coordinate (#2o);
  \draw(#2.north) ++(-2,0) coordinate (#2i0);
  \draw(#2.north) ++(2,0) coordinate (#2i1);
  \draw[hwbus](#2.east) ++(2,0) coordinate (#2cbit);
  \draw[hwwire, ->]  (#2cbit) -- (#2.east);
  \draw(#2cbit) node[right,blue] {#3};
}
\newcommand{\add}[2]{ % position, name
  \draw(#1) node(#2)[hwblock, minimum width=6ex, minimum height=3ex]{$+$};
	\ifthenelse{\boolean{tikzdebug}}{  \draw(#2.south west) ++(0,.4)node[blue,fill=yellow]{\tiny #2}; }{}
  \draw(#2.south) coordinate (#2o);
  \draw(#2.north) ++(-2,0) coordinate (#2i0);
  \draw(#2.north) ++(2,0) coordinate (#2i1);
}
\newcommand{\sub}[2]{ % position, name
  \draw(#1) node(#2)[hwblock, minimum width=6ex, minimum height=3ex]{$-$};
	\ifthenelse{\boolean{tikzdebug}}{  \draw(#2.south west) ++(0,.4)node[blue,fill=yellow]{\tiny #2}; }{}
  \draw(#2.south) coordinate (#2o);
  \draw(#2.north) ++(-2,0) coordinate (#2i0);
  \draw(#2.north) ++(2,0) coordinate (#2i1);
}
\newcommand{\addsub}[2]{ % position, name
  \draw(#1) node(#2)[hwblock, minimum width=6ex, minimum height=3ex]{$\pm$};
	\ifthenelse{\boolean{tikzdebug}}{  \draw(#2.south west) ++(0,.4)node[blue,fill=yellow]{\tiny #2}; }{}
  \draw(#2.south) coordinate (#2o);
  \draw(#2.north) ++(-2,0) coordinate (#2i0);
  \draw(#2.north) ++(2,0) coordinate (#2i1);
}
\else % Round adders for the poster
\newcommand{\add}[2]{ % position, name
  \draw(#1) node(#2)[hwblock, circle, minimum width=4ex]{$+$};
	\ifthenelse{\boolean{tikzdebug}}{  \draw(#2.south west) ++(0,.4)node[blue,fill=yellow]{\tiny #2}; }{}
  \draw(#2.south) coordinate (#2o);
  \draw(#2.135)  coordinate (#2i0);
  \draw(#2.45)  coordinate (#2i1);
  \draw(#2.north) ++(2,0) coordinate (#2i1);
}
\newcommand{\sub}[2]{ % position    \shift{#2oshift}{#5};1);
  \draw (#2i1) ++(1,0) node{\footnotesize $-$};
}
\newcommand{\addsub}[2]{ % position    \shift{#2oshift}{#5};1);
  \draw (#2i1) ++(1,0) node{\footnotesize $\pm$};
}
  \fi

\newcommand{\muxtwo}[4]{ % position, name, configbitname, color
  \draw(#1) node(#2)[hwblock, trapezium, trapezium stretches body, shape border rotate=180, minimum width=6ex, minimum height=3ex]{};
  \draw(#2.south) coordinate (#2o);
  \draw(#2.north) ++(-2,0) coordinate (#2i0) ++ (0,-1) node{\tiny 0};
  \draw(#2.north) ++(2,0) coordinate (#2i1)  ++ (0,-1) node{\tiny 1};
  \draw[hwbus](#2.east) ++(2,0) coordinate (#2cbit);
  \draw[hwwire, ->]  (#2cbit) -- (#2.east);
  \ifthenelse{\equal{#4}{0}}
  {
    \draw(#2cbit) node[red, right] {#3};
  }
  {
    \draw(#2cbit) node[#4, right] {#3};
  }
}

\newboolean{tikzdebug} % to show the names of each block
\setboolean{tikzdebug}{false}

\newcommand{\addnode}[5]{ % position, name, leftshift, rightshift, outputshift
  \add{#1}{#2}
  \nodedrawio{#1}{#2}{#3}{#4}{#5}{0}
}

\newcommand{\subnode}[5]{ % position, name, leftshift, rightshift, outputshift
  \sub{#1}{#2}
  \nodedrawio{#1}{#2}{#3}{#4}{#5}{0}
}

\newcommand{\muxtwonode}[6]{ % position, name, configbitname, shitf0, shift1, shift on the side, color
  \muxtwo{#1}{#2}{#3}{#6}
  \nodedrawio{#1}{#2}{#4}{#5}{0}{1}
	\ifthenelse{\boolean{tikzdebug}}{  \draw(#2.south west) ++(-1,.4)node[blue,fill=yellow]{\tiny #2}; }{}
}

\newcommand{\muxthreenode}[7]{ % position, name, configbitname, shitf0, shift1, shift2, shift on the side, color
  \draw(#1) node(#2)[hwblock, trapezium, trapezium stretches body, shape border rotate=180, minimum width=10ex, minimum height=3ex]{};
  \draw[hwbus] (#2.south) -- ++(0,-\nodeoutputwirelength) coordinate (#2o); % defining the coordinate;
  \draw(#2.north) ++(-4,0) coordinate (#2i0) ++ (0,-1) node{\tiny 0};
  \draw(#2.north) ++(0,0) coordinate (#2i1) ++ (0,-1) node{\tiny 1};
  \draw(#2.north) ++(4,0) coordinate (#2i2) ++ (0,-1) node{\tiny 2};
  \draw[hwbus](#2.east) ++(2,0) coordinate (#2cbit);
  \draw[hwwire, ->]  (#2cbit) -- (#2.east);
  \ifthenelse{\equal{#7}{0}}
  {
    \draw(#2cbit) node[red, right] {#3};
  }
  {
    \draw(#2cbit) node[#7, right] {#3};
  }
  \foreach \i in {0, 1, 2} {
    \draw[hwbus, <-](#2i\i) -- ++(0,\nodeinputwirelength) coordinate[pos=\nodeinputshiftpos](#2shift\i) coordinate (#2i\i);
  }
  \ifthenelse{\equal{#4}{0}}{}{\shiftLeft{#2shift0}{#4}; }
  \ifthenelse{\equal{#5}{0}}{}{\shift{#2shift1}{#5}; }
  \ifthenelse{\equal{#6}{0}}{}{\shiftRight{#2shift2}{#6}; }
}

\newcommand{\muxfournode}[8]{ % position, na29me, configbitname,configbitname , shift0, shift1, shift2, shift3
  \draw(#1) node(#2)[hwblock, trapezium, trapezium stretches body, shape border rotate=180, minimum width=14ex, minimum height=3ex]{};
  \draw[hwbus] (#2.south) -- ++(0,-\nodeoutputwirelength) coordinate (#2o); % defining the coordinate;

  \iffalse {
      \draw[hwwire, <-]  (#2.5) -- ++(3,0) coordinate (#2cbit) ;
      \draw(#2cbit) node[red, right] {#3};
      \draw[hwwire, <-]  (#2.-5) -- ++(2,0) coordinate (#2cbit) ;
      \draw(#2cbit) node[red, right] {#4};
    }
  \else {
      \draw[hwbus, <-]  (#2.east) -- ++(2,0) coordinate (#2cbit) ;
      \draw(#2cbit) node[red, right] {#4,#3};
    }
  \fi
  \foreach \i in {0, 1, 2, 3} {
    \draw(#2.north) ++(-6+4*\i,0) coordinate (#2i\i) ++ (0,-1) node{\tiny \i};
  }

  \ifthenelse{\equal{#5}{0} \AND \equal{#6}{0} \AND \equal{#7}{0}  \AND \equal{#8}{0} } {
    \foreach \i in {0, 1, 2, 3} {
      \draw[hwbus, <-](#2i\i) -- ++ (0,2) coordinate (#2i\i); % renaming the coordinate;
     }
  }{
    \foreach \i in {0, 1, 2, 3} {
      \draw[hwbus, <-](#2i\i) -- ++(0,\nodeinputwirelength) coordinate[pos=\nodeinputshiftpos](#2shift\i) coordinate (#2i\i);
    }
    \ifthenelse{\equal{#5}{0}}{}{\shift{#2shift0}{#5}; }
    \ifthenelse{\equal{#6}{0}}{}{\shift{#2shift1}{#6}; }
    \ifthenelse{\equal{#7}{0}}{}{\shift{#2shift2}{#7}; }
    \ifthenelse{\equal{#8}{0}}{}{\shift{#2shift3}{#8}; }
  }
}
\usepackage{multirow}
\usepackage{booktabs}
\usepackage{xcolor}
\usepackage{pgf}
\usepackage{graphicx}
\newcommand{\floor}[1]{\left\lfloor #1 \right\rfloor}

\renewcommand{\th}{^\text{th}}
\newboolean{anonymous}
\setboolean{anonymous}{true}   

\pgfplotsset{compat=1.18}

\usepgfplotslibrary{statistics}
\pgfplotsset{compat=1.18}

\usepackage{scalerel}
\usetikzlibrary{svg.path}

\definecolor{orcidlogocol}{HTML}{A6CE39}
\tikzset{
  orcidlogo/.pic={
    \fill[orcidlogocol] svg{M256,128c0,70.7-57.3,128-128,128C57.3,256,0,198.7,0,128C0,57.3,57.3,0,128,0C198.7,0,256,57.3,256,128z};
    \fill[white] svg{M86.3,186.2H70.9V79.1h15.4v48.4V186.2z}
                 svg{M108.9,79.1h41.6c39.6,0,57,28.3,57,53.6c0,27.5-21.5,53.6-56.8,53.6h-41.8V79.1z M124.3,172.4h24.5c34.9,0,42.9-26.5,42.9-39.7c0-21.5-13.7-39.7-43.7-39.7h-23.7V172.4z}
                 svg{M88.7,56.8c0,5.5-4.5,10.1-10.1,10.1c-5.6,0-10.1-4.6-10.1-10.1c0-5.6,4.5-10.1,10.1-10.1C84.2,46.7,88.7,51.3,88.7,56.8z};
  }
}

\newcommand\orcidicon[1]{\href{https://orcid.org/#1}{\mbox{\scalerel*{
\begin{tikzpicture}[yscale=-1,transform shape]
\pic{orcidlogo};
\end{tikzpicture}
}{|}}}}

\begin{document}
% paper title
% Titles are generally capitalized except for words such as a, an, and, as,
% at, but, by, for, in, nor, of, on, or, the, to and up, which are usually
% not capitalized unless they are the first or last word of the title.
% Linebreaks \\ can be used within to get better formatting as desired.
% Do not put math or special symbols in the title.
\title{Decompose, Optimize, and Reconstruct: Very Large Constant Multiplication at Scale}

% author names and affiliations
% use a multiple column layout for up to three different affiliations
%% AUTHORS
%\ifthenelse{\boolean{anonymous}}
%{ % anonymous=true START
%\author{\IEEEauthorblockN{Anonymous Authors}}
%} % anonymous=true END
{ % anonymous=false START
%\author{Hidden due to double-blind}
\author{
    \IEEEauthorblockN{
        Théo Cantaloube\textsuperscript{\orcidicon{0009-0007-3764-8301}}
    }
    \IEEEauthorblockA{
        INSA Lyon, CITI, UR3720\\
        Villeurbanne, France\\
        theo.cantaloube@insa-lyon.fr
    }
    \and
    \IEEEauthorblockN{
        Nicolai Fiege\textsuperscript{\orcidicon{0000-0002-4357-2119}}
    }
    \IEEEauthorblockA{
        University of Kassel\\
        Kassel, Germany
    }
    \and
    \IEEEauthorblockN{
        Anastasia Volkova, Christine Solnon
    }
    \IEEEauthorblockA{
        Inria, INSA Lyon, CITI, UR3720\\
        Villeurbanne, France
    }
}

% anonymous=false END

% conference papers do not typically use \thanks and this command
% is locked out in conference mode. If really needed, such as for
% the acknowledgment of grants, issue a \IEEEoverridecommandlockouts
% after \documentclass

% for over three affiliations, or if they all won't fit within the width
% of the page, use this alternative format:
% 
%\author{\IEEEauthorblockN{Michael Shell\IEEEauthorrefmark{1},
%Homer Simpson\IEEEauthorrefmark{2},
%James Kirk\IEEEauthorrefmark{3}, 
%Montgomery Scott\IEEEauthorrefmark{3} and
%Eldon Tyrell\IEEEauthorrefmark{4}}
%\IEEEauthorblockA{\IEEEauthorrefmark{1}School of Electrical and Computer Engineering\\
%Georgia Institute of Technology,
%Atlanta, Georgia 30332--0250\\ Email: see http://www.michaelshell.org/contact.html}
%\IEEEauthorblockA{\IEEEauthorrefmark{2}Twentieth Century Fox, Springfield, USA\\
%Email: homer@thesimpsons.com}
%\IEEEauthorblockA{\IEEEauthorrefmark{3}Starfleet Academy, San Francisco, California 96678-2391\\
%Telephone: (800) 555--1212, Fax: (888) 555--1212}
%\IEEEauthorblockA{\IEEEauthorrefmark{4}Tyrell Inc., 123 Replicant Street, Los Angeles, California 90210--4321}}

% use for special paper notices
%\IEEEspecialpapernotice{(Invited Paper)}

\maketitle

% As a general rule, do not put math, special symbols or citations in the abstract
\begin{abstract}
    Efficient arithmetic circuit design for resource-constrained hardware involves challenging combinatorial optimization problems, among which Multiple Constant Multiplication (MCM) is a prominent example.
    MCM aims at implementing multiplications by fixed integer constants using bit-shifts and additions/subtractions but optimal methods are typically limited to moderately-sized constants, e.g. 12 bits.
    For practical applications targeting larger precision, Very large Constant Multiplication (VLCM) is solved instead. 
    Existing approaches typically address VLCM through a heuristic flow that decomposes large constants into patterns, applies MCM optimization techniques on moderately-sized targets, and reconstructs the final result.
    This paper proposes multiple improvements to this flow: new declarative optimization models for the pattern selection and for the reconstruction, as well as applying recent optimal MCM models. 
    The cornerstones of the obtained improvements are (i) allowing the patterns to overlap, minimising the number of unique target constants for the MCM step and (ii) performing the reconstruction step optimally, instead of heuristically. 
    In addition, we propose a globally-optimal VLCM approach and characterize its limits. 
    We employ a mix of constraint programming and SAT to solve each step.
    Experimental results on synthetic and real-life signal processing and cryptographic benchmarks, with coefficient word lengths ranging from tens to thousands of bits, demonstrate that the proposed approach scales to very large precisions and consistently outperforms existing baselines.
\end{abstract}

% no keywords here
% keywords in submission form:

% For peer review papers, you can put extra information on the cover
% page as needed:
% \ifCLASSOPTIONpeerreview
% \begin{center} \bfseries EDICS Category: 3-BBND \end{center}
% \fi
%
% For peerreview papers, this IEEEtran command inserts a page break and
% creates the second title. It will be ignored for other modes.
\IEEEpeerreviewmaketitle

\section{Introduction}

The Multiple Constant Multiplication (MCM) problem is a well-known combinatorial optimization problem with numerous applications, including deep neural networks~\cite{HardieckHWMKZ23, garcia_mcm_dnn_2025, HabermannKKV22}, discrete transforms~\cite{GarridoM21}, and digital signal processing~\cite{johansson_bit-level_2007, AksoyFM14, mcm_iir_2022,Volkova0DK23}.
Its relevance stems from the optimization of multiplications when the target constants are known \textit{a priori} which permits to build custom constant multipliers. 

MCM typically aims at finding a rewriting of multiplications by known integers using bit-shifts and additions/subtractions into so-called shift-and-add graphs. For example, $11X = X \ll 4 - X - X \ll 2$ as illustrated in Fig.~\ref{fig:intro_ag}(a). 
Bit-shifts are essentially implemented as wiring in hardware whose cost is too unpredictable, hence the implementation cost is usually measured by the number of adders/subtractors. 
Although this provides a reasonable proxy for hardware cost, more fine-grained metrics, that will not be studied in the paper, are sometimes considered in the literature~\cite{johansson_detailed_2005, aksoy_optimization_2007, lou_fine-grained_2015, garcia23, fiege24}.

There are multiple approaches to solve MCM \cite{Dempster1994, dempster_use_1995, voronenko_multiplierless_2007, aksoy_search_2010, kumm_optimal_2018, garcia23, fiege24, cantaloube25} and it has been conjectured to be $\cal NP$-hard \cite{Cappello1984}. Indeed, when there are several target constants, difficulty lies in determining which and how intermediate values should be shared across the different targets to minimize the total number of adders. For example, if the target constants are $93$ and $11$, simple combining of graphs yields a 4-adder solution but an optimal graph, passing through a different set of intermediate constants (called fundamentals), will need only 3, as illustrated in Fig.~\ref{fig:intro_ag}. 

To solve $\cal NP$-hard combinatorial optimization problems like MCM, a first option is to develop dedicated approaches, \textit{e.g.} ~\cite{aksoy_search_2010}. However, obtaining optimal solutions within reasonable time is generally challenging: it requires sophisticated bounding functions and heuristics to limit combinatorial explosion, as well as careful implementation with efficient data structures. Beyond being time-consuming, this implementation phase is prone to errors, and verifying correctness or optimality can be difficult. Reproducibility is also a concern, since adapting such algorithms to related problems with both commonalities and differences may again prove complex and costly.
Declarative approaches (\textit{e.g.}, Integer Linear Programming (ILP), Boolean SATisfiability (SAT), and Constraint Programming (CP)) offer an attractive alternative, as they only require the design of a mathematical model that can be solved by off-the-shelf solvers.

\begin{figure}[t]
    \centering
    \begin{subfigure}[t]{0.30\linewidth}
        \centering
        \scalebox{0.7}{%
    \begin{tikzpicture}[scale=0.2]
        \subnode{-2,-7}{sub0}{4}{0}{0}
        \subnode{2,-16}{sub1}{0}{2}{0}
        \draw (0,3) node(x) {$1$};
        \draw[hwbus] (x) -- ++(0,-3) coordinate (x);
        
        \draw (sub0o) ++(0,-1.5) node {$15$};
        \draw[hwbus, ->] (sub1o) -- ++(0,-2) node[right] {$\textbf{11}$};
        
        \draw[hwbus] (x.south) -| (sub0i0);
        \draw[hwbus] (x.south) -| (sub0i1);
        \draw[hwbus] (sub0o) -| (sub1i0);
        \draw[hwbus] (x.south) -| (sub1i1);
    \end{tikzpicture}
}
        \caption{SCM for 11}
        \label{fig:subA}
    \end{subfigure}
    \hfill
    \begin{subfigure}[t]{0.30\linewidth}
        \centering
        \scalebox{0.7}{%
    \begin{tikzpicture}[scale=0.2]
        \subnode{0,-7}{sub0}{5}{0}{0}
        \addnode{0,-16}{add1}{2}{0}{0}
        \draw (0,3) node(x) {$1$};
        \draw[hwbus] (x) -- ++(0,-3) coordinate (x);
        
        \draw (sub0o) ++(0,-1.5) node {$31$};
        \draw[hwbus, ->] (add1o) -- ++(0,-2) node[right] {$\textbf{93}$};
        
        \draw[hwbus] (x.south) -| (sub0i0);
        \draw[hwbus] (x.south) -| (sub0i1);
        \draw[hwbus] (sub0o) -| (add1i0);
        \draw[hwbus] (sub0o) -| (add1i1);
    \end{tikzpicture}
}

%3 = (1 \ll 2) - 1 \; \quad 35 = (1 \ll 5) + 3 \; \quad 
%93 = (1 \ll 7) - 35
        \caption{SCM for 93}
        \label{fig:subB}
    \end{subfigure}
    \hfill
    \begin{subfigure}[t]{0.30\linewidth}
        \centering
        \scalebox{0.7}{%
    \begin{tikzpicture}[scale=0.2]
        \addnode{0,-7}{add0}{1}{0}{0}
        \addnode{-5,-16}{add1}{3}{0}{0}
        \subnode{5,-16}{sub2}{5}{0}{0}
        \draw (0,3) node(x) {$1$};
        \draw[hwbus] (x) -- ++(0,-3) coordinate (x);
        
        \draw (add0o) ++(0,-1.5) node {$3$};
        \draw[hwbus, ->] (add1o) -- ++(0,-2) node[right] {$\textbf{11}$};
        \draw[hwbus, ->] (sub2o) -- ++(0,-2) node[right] {$\textbf{93}$};
        
        \draw[hwbus] (x.south) -| (add0i0);
        \draw[hwbus] (x.south) -| (add0i1);
        \draw[hwbus] (x.south) -| (add1i0);
        \draw[hwbus] (add0o) -| (add1i1);
        \draw[hwbus] (add0o) -| (sub2i0);
        \draw[hwbus] (add0o) -| (sub2i1);
    \end{tikzpicture}
}

%3 = (1 \ll 1) + 1 \; \quad 11 = (1 \ll 3) + 3 \; \quad 93 = (3 \ll 5) - 3.
        \caption{MCM $\{11, 93\}$}
        \label{fig:subC}
    \end{subfigure}
    \caption{Shift-and-add graphs: blue arrows denote left shifts, below each adder is the result of the addition.}
    \label{fig:intro_ag}
\end{figure}

MCM has recently been optimally solved with ILP \cite{garcia23}, SAT \cite{fiege24}, and CP \cite{cantaloube25}. 
A recent comparison~\cite{cantaloube25} demonstrated that SAT and CP outperform ILP. While scaling well with increasing number of constants (up to 50 in the considered benchmark), these approaches struggle with coefficient size larger than 12 bits and can rarely go beyond 16 bits in a reasonable time.

Yet, for some applications the width may be much larger, \textit{e.g.}, around 50 bits for digital signal processing, and even 500 bits for cryptography. In this case, the MCM problem is called the Very Large Constant Multiplication (VLCM) problem. 

Various approaches of the large scale multiplications have been proposed in the literature.
In \cite{Rafferty2017}, a hybrid approach combining Comba and Karatsuba multipliers is studied; however, both operands are treated as variables, whereas in VLCM the target constants are fixed.
Residue Number System (RNS) large-scale multiplication schemes \cite{Chaves2007} were also considered, but the conversion overhead required to translate operands into the RNS representation incurs prohibitive costs in our setting.

The first approach for shift-and-add VLCM has been introduced in \cite{aksoy_multiplierless_2022}.
Given a set $T$ of base-2 target constants, this approach proceeds in 3 steps:
\begin{enumerate}
    \item[1] \textbf{Pattern decomposition:} $T$ is decomposed into a multi-set $M$ of base-2 smaller patterns (of size at most $w$ bits) so that each 1 in a target constant of $T$ is covered by a 1 of exactly one pattern of $M$; 
    \item[2] \textbf{Optimization for unique patterns}: the MCM problem is solved for the set of target constants $S$ obtained from $M$ by keeping \textit{unique patterns};
    %CS: I proposed to remove "that are moderately-sized;" because the size of these patterns has not changed between step 1 and step 2.
    \item[3] \textbf{Reconstruction}: The initial large target constants of $T$ are reconstructed from the outputs of the Step 2 MCM.
\end{enumerate}

\begin{figure}[t]
    \centering
    \scalebox{0.7}{%
    \begin{tikzpicture}[scale=0.2]
        \addnode{-5,-7}{add0}{1}{0}{0}
        \addnode{ 5,-7}{add1}{2}{0}{0}
        \addnode{ 0,-19}{add2}{0}{3}{0}
        \addnode{ 0,-28}{add3}{6}{0}{0}
        
        \draw (0,3) node(x) {$1$};
        \draw[hwbus] (x) -- ++(0,-3) coordinate (x);
        
        \draw (add0o) ++(0,-2) node {$11_2$};
        \draw (add1o) ++( 0,-2) node {$101_2$};
        \draw (add2o) ++( 6, 0) node {$101011_2$};
        \draw[hwbus, ->] (add3o) -- ++(0,-2) node[right] {$\textbf{101011101011}_2$};
        
        \draw[hwbus] (x.south) -| (add0i0);
        \draw[hwbus] (x.south) -| (add0i1);
        \draw[hwbus] (x.south) -| (add1i0);
        \draw[hwbus] (x.south) -| (add1i1);
        \draw[hwbus] (add0o) -| (add2i0);
        \draw[hwbus] (add1o) -| (add2i1);
        \draw[hwbus] (add2o) -| (add3i0);
        \draw[hwbus] (add2o) -| (add3i1);
        
        \node[anchor=west] at (-28, -7)  {MCM for unique patterns};
        \draw[densely dotted] (-28,-9) -- (12,-9);
        \node[anchor=west] at (-28, -11) {Unique patterns $S$};
        \draw[densely dotted] (-28,-13) -- (12,-13);
        \node[anchor=west] at (-28, -15) {Reconstruction};
    \end{tikzpicture}
}
    \caption{Adder-graph solution for a three-step approach on a toy example $2795=101011101011_2$. Using a regular 3-bit decomposition, the pattern multiset is $M=\{\!\{101,011,101,011\}\!\}$. The unique pattern set $S=\{101, 011\}$ is computed with two adders (top), reconstruction uses CSE and requires two adders (bottom).}
    \label{fig:3_step_intro}
\end{figure}

In \cite{aksoy_multiplierless_2022}, Step 1 (pattern decomposition) divides the target constants into regular equal-sized chunks, typically of 4, 8, or 12 bits.
Sequences of $r$ consecutive 0s, where $r \mod w = 0$, are skipped.
Similarly, sequences of $r$ consecutive 1s are replaced by $2^r - 1$.
Step 2 (optimization for unique patterns) is implemented with a dedicated exact/heuristic solver~\cite{aksoy_search_2010} but could be solved with any existing MCM heuristic, \textit{e.g.} H$_\text{cub}$ \cite{voronenko_multiplierless_2007} or RPAG~\cite{kumm12}, or optimal MCM approaches~\cite{garcia23, fiege24, cantaloube25}.
Finally, the reconstruction in Step 3 is based on a heuristic using Common Sub-expression Elimination (CSE) \cite{hartley_subexpression_1996}.
Figure~\ref{fig:3_step_intro} illustrates a minimalist example of the approach.
The overall approach~\cite{aksoy_multiplierless_2022} is a part of a dedicated tool called TOLL and also contains additional optimizations to reduce the gate-level area and the delay of the shift-adds design at the cost of a small overhead. 
More recently, the tool LEIGER targeting VLCM for ASICs was introduced in \cite{aksoy_multiplierless_2024}, and incorporates many new architectural features but the pattern decomposition approach remains the same. It supports multiple architectural paradigms, including optimized carry-save adder reduction, hybrid 2/3-input operator schemes, and compressor-tree-based structures and enabling systematic exploration of area–delay trade-offs.

In this paper, we use the term \textit{overlapping} patterns to refer to binary patterns that overlap without conflicting; that is, whenever a bit is set to 1 in one pattern, the corresponding bit in the other pattern must be 0. Unfortunately, \cite{Thong2009} used the same terminology to describe what we consider more appropriately termed \textit{clashing} patterns, namely patterns for which a bit set to 1 in one pattern coincides with a bit set to 1 in the other. Despite the similar terminology, the two notions are fundamentally different and should not be confused.

\subsection*{Overview and Contributions of the Paper}

In \cite{aksoy_multiplierless_2022}, the patterns in Step 1 must be disjoint.
In Section~\ref{sec:step(i)SAT}, we introduce a \textbf{new pattern decomposition} approach for Step 1 where patterns have an upper bound $w$ on size and \textbf{may overlap}, thus allowing us to reduce the number of unique patterns. We introduce a SAT model for computing such an overlapping decomposition that lexicographically minimizes (i) the total number of patterns in $M$ and (ii) the number of unique patterns in $S$ for a given pattern width. 

In Section \ref{sec:rec}, we \textbf{improve the reconstruction} on Step 3 by extending the SAT model~\cite{fiege24} for MCM. 
This permits us to obtain a 3-step methodology that solves each task optimally based on SAT models.
As we shall exhibit, this is also very time-consuming.
Hence, we introduce in Section~\ref{sec:CSDh} an efficient \textbf{heuristic approach} based on Canonic Sign Digit (CSD) representation of the target constants.

Finally, in Section \ref{sec:CSDe}, we explore the feasability of a globally optimal VLCM solution and introduce \textbf{the first exact approach} for the 3-step VLCM. 
We propose to enumerate all possible decompositions at Step~1, where patterns may both have variable widths and overlap.
For each decomposition, we optimally solve Step 2 (with either the CP approach of \cite{cantaloube25} or the SAT approach of \cite{fiege24}) and Step 3 (with the SAT approach introduced in Section~\ref{sec:rec}).

In Section~\ref{sec:experiments}, we experimentally show that hardware cost reductions can be obtained by (i) using CSD for pattern decomposition, and minimizing the total pattern count with overlap; and (ii) using MCM also for the reconstruction.

All of our contributions are summed up in Fig.~\ref{fig:overall}.
\begin{figure*}[t]
    \centering
    \begin{tikzpicture}[scale=0.17]
    \node (in) at (0,12) {$T$};
    \node[draw, align=center, minimum size=7mm] (D) at (0,6) {(1) Divide};
    \node (M) at (0,0) {Multiset of patterns $M$};
    \node[align=center] (S) at (0,-7) {Set of unique\\patterns $S$};
    \node[draw, align=center, minimum size=7mm] (O) at (-10,-14) {(2) Optimal MCM\\for patterns};
    \node[draw, align=center, minimum size=10mm] (R) at (10,-14) {(3) Reconstruct of\\target constants};
    \node (Oo) at (-10,-21) {$AG_{1 \rightarrow S}$};
    \node (Ro) at (10,-21) {$AG_{S \rightarrow T}$};
    \node (F) at (0,-25) {$AG_{1 \rightarrow T}$};
    \draw[->] (in.south) -- (D.north);
    \draw[->] (D.south) -- (M.north);
    \draw[->] (M.south) -- (S.north);
    \draw[->] (S.west) -| (O.north);
    \draw[->] (S.east) -| (R.north);
    \draw[->] (O.south) -- (Oo.north);
    \draw[->] (R.south) -- (Ro.north);
    \draw[->] (Oo.east) -| (F.north);
    \draw[->] (Ro.west) -| (F.north);

    \node[anchor=west] (D1) at ($(D.east) + (4,6)$) {Chunk Divide \cite{aksoy_multiplierless_2022}};
    \node[anchor=west, align=left] (D2) at ($(D.east) + (4,2)$) {\textbf{Min. number of patterns} \\ \textbf{(SAT, proposed)}};
    \node[anchor=west] (D3) at ($(D.east) + (4,-2)$) {\textbf{Optimal (enumeration, proposed)}};
    \draw[->] (D.east) -- (D1.west);
    \draw[->] (D.east) -- (D2.west);
    \draw[->] (D.east) -- (D3.west);
    \node[anchor=east, align=right] (O1) at ($(O.west) + (-4,2)$) {Heuristic MCM (RPAG \cite{kumm12})};
    \node[anchor=east, align=right] (O2) at ($(O.west) + (-4,-2.5)$) {Optimal MCM \\ (CP \cite{cantaloube25} / SAT \cite{fiege24} / BnB \cite{aksoy_search_2010})};
    \draw[->] (O.west) -- (O1.east);
    \draw[->] (O.west) -- (O2.east);
    \node[anchor=west, align=left] (R1) at ($(R.east) + (4,2)$) {CSE + Lineq \cite{aksoy_multiplierless_2022, aksoy_multiplierless_2024}};
    \node[anchor=west, align=left] (R2) at ($(R.east) + (4,-2.5)$) {\textbf{Modified optimal MCM} \\ \textbf{(SAT, proposed)}};
    \draw[->] (R.east) -- (R1.west);
    \draw[->] (R.east) -- (R2.west);
\end{tikzpicture}
    \caption{Overview of the methodology and contributions}
    \label{fig:overall}
\end{figure*}

\section{New SAT models for Pattern Decomposition and Target Reconstruction}

\subsection{Pattern Decomposition}
\label{sec:step(i)SAT}

In \cite{aksoy_multiplierless_2022}, the underlying modeling assumes that patterns are disjoint.
The first contribution of this paper is the introduction of \textbf{overlapping patterns} that allows us to reduce the number of unique patterns in $S$, as illustrated in Fig.~\ref{fig:subconstant}.

\begin{figure}[h!t!]
    \centering
    \definecolor{pastelgreen}{RGB}{187,220,188}
\definecolor{pastelorange}{RGB}{255,217,179}
\definecolor{pastelred}{RGB}{255,179,186}
\definecolor{pastelcyan}{RGB}{186,225,255}
\definecolor{pastelyellow}{RGB}{255,255,186}
\definecolor{pastelblue}{RGB}{186,201,255}

\begin{tikzpicture}[scale=0.4]
    \newcommand{\bit}[2][]{    % Macro : dessine un bit et avance
        \draw[fill=#1] (cur) rectangle ++(1,1);
        \node at ($(cur)+(0.5,0.5)$) {#2};
        \coordinate (cur) at ($(cur)+(1,0)$);
    }

    \begin{scope}[yshift=0.5cm]
        \coordinate (cur) at (0,0);
        \bit[white]{1}
        \bit[white]{1}
        \bit[white]{1}
        \bit[white]{0}
        \bit[white]{1}
        \bit[white]{0}
        \bit[white]{1}
        \bit[white]{1}
        \bit[white]{1}
        \bit[white]{1}
        \bit[white]{0}
        \bit[white]{1}
        \bit[white]{0}
        \bit[white]{1}
        \bit[white]{1}
        \bit[white]{1}
        \draw[dashed] (-1,-0.9) -- (16,-0.9);
    \end{scope}

    \begin{scope}[yshift=-2.3cm]
        \node[rotate=90, anchor=south] at (-2.6,0.5) {chunk};
        \node[rotate=90, anchor=south] at (-1.6,0.5) {divide};

        \coordinate (cur) at (-1,0);
        \bit[pastelyellow]{0}
        \bit[pastelyellow]{1}
        \bit[pastelyellow]{1}
        \bit[pastelyellow]{1}

        \coordinate (cur) at (4,0);
        \bit[pastelred]{1}
        \bit[pastelred]{0}
        \bit[pastelred]{1}
        \bit[pastelred]{1}

        \coordinate (cur) at (8,0);
        \bit[pastelcyan]{1}
        \bit[pastelcyan]{1}
        \bit[pastelcyan]{0}
        \bit[pastelcyan]{1}

        \coordinate (cur) at (12,0);
        \bit[pastelyellow]{0}
        \bit[pastelyellow]{1}
        \bit[pastelyellow]{1}
        \bit[pastelyellow]{1}

        \draw[dashed] (-1,-0.9) -- (16,-0.9);
    \end{scope}

    \begin{scope}[yshift=-5.1cm]
        \node[rotate=90, anchor=south] at (-2.6, 0) {over-};
        \node[rotate=90, anchor=south] at (-1.6, 0) {lapping};

        \coordinate (cur) at (-1,0);
        \bit[pastelyellow]{0}
        \bit[pastelyellow]{1}
        \bit[pastelyellow]{1}
        \bit[pastelyellow]{1}
        
        \coordinate (cur) at (3.9,-1.1);
        \bit[pastelblue]{1}
        \bit[pastelblue]{0}
        \bit[pastelblue]{0}
        \bit[pastelblue]{1}

        \coordinate (cur) at (6,0);
        \bit[pastelblue]{1}
        \bit[pastelblue]{0}
        \bit[pastelblue]{0}
        \bit[pastelblue]{1}

        \coordinate (cur) at (8.1,-1.1);
        \bit[pastelblue]{1}
        \bit[pastelblue]{0}
        \bit[pastelblue]{0}
        \bit[pastelblue]{1}

        \coordinate (cur) at (12,0);
        \bit[pastelyellow]{0}
        \bit[pastelyellow]{1}
        \bit[pastelyellow]{1}
        \bit[pastelyellow]{1}
    \end{scope}
\end{tikzpicture}
    \caption{Subdividing $60375_{10} = 1110101111010111_2$ into patterns of size $w=4$.
    Chunk divide skips zeros between patterns leading to $M=\{\!\{111,1011,1101,111\}\!\}$ and $S=\{111,1011,1101\}$.
    Overlapping leads to $M=\{\!\{111,1001,1001,1001,111\}\!\}$ and $S=\{111,1001\}$.}
    \label{fig:subconstant}
\end{figure}

Therefore, we aim to optimally solve the pattern decomposition problem under a given metric.
Possible metrics include minimizing the number of unique patterns in $S$ (in Fig.~\ref{fig:subconstant} overlapping, there are 2 patterns in $S$) or minimizing the total number of patterns in the multiset $M$ (in Fig.~\ref{fig:subconstant} overlapping, there are 5 patterns in $M$).
However, minimizing the number of unique patterns systematically yields the pattern set $S=\{1\}$, whose use for reconstructing the target constants would incur a very high computational cost.
On the other hand, minimizing the number of total patterns in the multiset $M$ would certainly yield chunk divide pattern decomposition.
Consequently, we instead aim to minimize the total number of patterns as first objective and the number of unique patterns as second objective, expecting the resulting pattern set $S$ to remain suitable for efficient reconstruction.

As an example consider the target constant $15 = 1111_2$ with pattern size $w=3$.
The minimum total number of patterns in $M$ is two, but there are multiple possible decompositions.
Dividing $15$ naively into 3-bit chunks gives $M=\{\!\{001,111\}\!\}$ with two unique patterns.
Alternatives with only one unique pattern are $M=\{\!\{011,011\}\!\}$ and $M=\{\!\{101,101\}\!\}$.

To that extent, let us present a SAT-based pattern decomposition model built for a given set $T$ of target constants, a given number $\#\,\text{TP}$ of total patterns as well as a given upper bound $\#\,\text{UP}^\text{max}$ of unique patterns with the following variables:
\begin{itemize}
    \item $\alpha^d_x$: Represents the $x\th$ bit of the $d\th$ unique pattern;
    \item $\beta^t_x$: Represents the $x\th$ bit of the $t\th$ pattern in $M$;
    \item $\gamma^{m,t}_{y,x}$: Indicates whether the $y\th$ bit of the $m\th$ target constant gets mapped to the $x\th$ bit of the $t\th$ pattern;
    \item $\delta^{d,t}$: Indicates whether the $t\th$ pattern is equal to the $d\th$ unique pattern;
    \item $\varepsilon^{m,t}$: Indicates whether the $t\th$ pattern receives bits from the $m\th$ target constant.
\end{itemize}
The index domains are given by 
$d \in \llbracket 1, \#\,\text{UP}^\text{max} \rrbracket$; 
$t \in \llbracket 1, \#\,\text{TP} \rrbracket$;
$x \in \llbracket 1, w \rrbracket$;
$m \in \llbracket 1, \#T \rrbracket$;
$y \in \llbracket 0, W-1 \rrbracket$.

Note that only those $\gamma^{m,t}_{y,x}$ are needed in the SAT formula, for which the $m\th$ target constant in $T$ has a one in its bit representation at bit position $w$.
The resulting SAT formula is built from several sub-formulae, each representing a high-level constraint:
\begin{itemize}
    \item [$\varphi_1$:] Each 1 in each target constant is mapped to exactly one pattern;
    \item [$\varphi_2$:] Patterns are properly defined according to their mappings;
    \item [$\varphi_3$:] Prohibit invalid combinations of ones for each pattern; %Only valid combinations of ones can be mapped to the same pattern;
    \item [$\varphi_4$:] Each pattern is equal to one unique pattern;
    \item [$\varphi_5$:] Each pattern can receive its bits only from one target constant;
    \item [$\varphi_6$:] Each unique pattern is odd.
\end{itemize}
The sub-formulae are defined as
\begingroup
    \allowdisplaybreaks  % allow page breaks, otherwise the layout leaves too much whitespace
    \begin{align*}
        \nonumber
        \varphi_1 :=& \, \left( \bigwedge_{m,y} \bigvee_{t,x} \gamma^{m,t}_{y,x} \right) \\
        \land& \, 
        \left( \bigwedge_{m,y,t_1,x_1} \bigwedge_{\substack{t_2,x_2: \\ t_2 \geq t_1, x_2 \geq x_1 \\ t_1 \neq t_2 \lor x_1 \neq x_2}} \neg \gamma^{m,t_1}_{y,x_1} \lor \neg \gamma^{m,t_2}_{y,x_2} \right) \\
        \varphi_2 := & \, 
        \left( \bigwedge_{m,t,x,y} \neg \gamma^{m,t}_{y,x} \lor \beta^t_x \right)
        \land
        \left( \bigwedge_{t,x} \neg \beta^t_x \lor \bigvee_{m,y} \gamma^{m,t}_{y,x} \right)\\
        \varphi_3 := & \, 
        \bigwedge_{m,t,x_1,y_1} \bigwedge_{\substack{x_2,y_2 \\ y_2 - y_1 \neq x_2 - x_1 \\ y_1 \leq y_2}} \neg \gamma^{m,t}_{y_1,x_1} \lor \neg \gamma^{m,t}_{y_2,x_2}\\
        \nonumber
        \varphi_4 := & \, 
        \left( \bigwedge_{t} \bigvee_{d} \delta^{d,t} \right) 
        \land \,
        \left( \bigwedge_{t,d,x} \alpha^d_x \lor \neg \beta^t_x \lor \neg \delta^{d,t} \right) \\
        \land & \,
        \left( \bigwedge_{t,d,x} \neg \alpha^d_x \lor \beta^t_x \lor \neg \delta^{d,t} \right) \\
        \varphi_5 := & \, 
        \left( \bigwedge_{m,t,x,y} \neg \gamma^{m,t}_{y,x} \lor \varepsilon^{m,t} \right)
        \land
        \left( \bigwedge_{\substack{t, m_1, m_2 \\ m_2 > m_1}} \neg \varepsilon^{m_1,t} \lor \neg \varepsilon^{m_2,t} \right) \\
        \varphi_6 := & \, 
        \bigwedge_{d} \alpha^d_0 .
    \end{align*}
\endgroup

The first part in $\varphi_1$ makes sure that each target constant is mapped to at least one pattern, while the second part prohibits mappings to more than one pattern.
Due to symmetry of the disjunction, we are able to leave out duplicate clauses that would arise in the second part.

The clauses in the first part of $\varphi_2$ implement the relationship \enquote{if the $w\th$ 1 of the bit representation of the $m\th$ target constant is mapped to the $x\th$ bit in the $t\th$ pattern, then that bit of this pattern must be set to 1}.
Similarly, the second part of $\varphi_2$ encodes \enquote{if not a single 1 gets mapped to the $x\th$ bit of the $t\th$ pattern, then that pattern must have a zero at this bit position}.

Sub-formula $\varphi_3$ ensures that only compatible bit positions are mapped to the same pattern.
As a counter-example consider that $\gamma^{m,t}_{0,0}$ and $\gamma^{m,t}_{2,1}$ would be set to \texttt{true}, simultaneously (\emph{i.e.}, $y_2 - y_1 = 2-0 \neq 1-0 = x_2 - x_1$). 
This would mean that bits 0 and 2 of the $m\th$ target constant are mapped to bit positions 0 and 1 of pattern $t$, which would represent an invalid mapping.

The first set of clauses in $\varphi_4$ forces each pattern to be equal to at least one unique pattern.
The clauses in the second and third parts of $\varphi_4$ implement the relationship \enquote{if pattern $t$ is equal to unique pattern $d$, then all their bits must be equal} (i.e., $\delta^{d,t} \implies (\alpha^d_x \Leftrightarrow \beta^t_x)$).

The sub-formula $\varphi_5$ connects the $\gamma^{m,t}_{w,x}$ variables with the $\varepsilon^{m,t}$ variables, and ensures that each pattern cannot simultaneously receive bits from different target constants.

Finally, $\varphi_6$ is used to limit the search space for the solver by only allowing odd patterns, since each even pattern can be represented as an odd pattern with a right shift.

All sub-formulae are given as CNFs, so we can construct the final formula via
\begin{equation}
    \varphi_\text{final} := \varphi_1 \land \varphi_2 \land \varphi_3 \land \varphi_4 \land \varphi_5 \land \varphi_6.
\end{equation}

A satisfying assignment to $\varphi_\text{final}$ either allows us to construct a partitioning of the target constants into $\#\,\text{TP}$ patterns with at most $\#\,\text{UP}^\text{max}$ unique patterns ($\varphi_\text{final}$ is \emph{satisfiable}), or it is guaranteed that such a partitioning does not exist ($\varphi_\text{final}$ is \emph{unsatisfiable}).

We perform an optimization by systematically varying the total number of patterns as well as the available unique patterns and repeatedly checking satisfiability of the resulting $\varphi_\text{final}$.

The number of unique patterns influences costs of the MCM part for pattern synthesis.
Generating many different target constants is, in general, more expensive than generating only few different targets (disregarding their individual complexities).
Furthermore, the resulting MCM instances become more difficult to solve, as MCM solving time increases with the number of targets.

The total number of patterns influences costs of the reconstruction part. 
Assembling the targets from a high number of individual patterns is, in general, more expensive than assembling them from only a few patterns (disregarding optimizations such as sub-expression elimination).

The minimal total number of patterns for a given pattern size $w$ can be obtained via the chunk divide approach shown in Fig.~\ref{fig:subconstant} (yielding the minimal value for $\#\,\text{TP}$).
The minimal number of \textit{unique} patterns can then be obtained by starting with $\#\,\text{UP}^\text{max} = 1$ and increasing the limit by one each time the solver reports \emph{unsatisfiability}, guaranteeing the first feasible solution to be optimal.

\subsection{Reconstruction of large target constants}
\label{sec:rec}

A naive approach to reconstruct target constants out of unique patterns computed with MCM, is simply to use shift-and-add to ``glue" the patterns in the correct order. 
An improvement proposed in \cite{aksoy_multiplierless_2022} is to use common subexpression elimination and linear equation optimization.
In this work, we target an optimal approach for this step using the MCM optimization. 
Compared to the classic MCM problem, which always takes as unique input of the adder-graph the value \enquote{1}, the VLCM reconstruction that we propose starts with all numeric values of the unique patterns.
To do so, we modified the SAT-based MCM solver \cite{fiege24} to accept a given pattern set as inputs but do not present the details due to lack of space.

\section{CSD-based heuristic}
\label{sec:CSDh}

The previously introduced subdivision and reconstruction steps based on SAT solvers are computationally expensive.
A promising intermediate approach, offering solution quality between that of SAT-based methods and \cite{aksoy_multiplierless_2022}, while retaining a runtime comparable to the latter, is to employ heuristics based on CSD representations of the target constants.

Signed Digit (SD) forms are ternary representations of integers whose digits take values in $\{1,0,\overline{1}\}$ (where $\overline{1}=-1$).
Canonical Signed Digit (CSD) forms are SD forms where there are no consecutive non-zero digits.
Although integers can have multiple SD representations, they have a unique CSD form.
CSD can be obtained in linear time w.r.t the word length and minimizes the number of non-zero digits among all SD forms \cite{Reitwiesner1960}.
For a given integer of word length $w$, the word length of its CSD form is in $\{ w, w+1 \}$.
As there are no consecutive non-zero digits, the quantity of non-zero digits in CSD is bounded by $\floor{\frac{w+1}{2}}$.
Consequently, CSD representations have been extensively used in the study of Single and Multiple Constant Multiplication problems \cite{Hartley1991, hartley_subexpression_1996, Dempster1994}, since they typically yield lower implementation costs while preserving polynomial-time complexity.

For our new CSD-based heuristic, the pattern decomposition part is realized with a chunk divide linear process, similar to \cite{aksoy_multiplierless_2022}, except that target constants are first encoded in CSD forms, and 0 bits are skipped so that patterns always start with a non-zero digit. When chunks are negative, \textit{i.e.}, the MSB is $\overline{1}$, we consider its absolute value in order to leave to the reconstruction step the choice of subtracting or not. In some cases, this reduces the cardinality of $S$, when both a number and its negation belong to $M$. Hence, we define $S$ as the set containing the absolute values of the patterns of $M$.
This process is illustrated in Fig.~\ref{fig:subdivision_csd}.

\begin{figure}[ht!]
    \centering
    \definecolor{pastelgreen}{RGB}{187,220,188}
\definecolor{pastelorange}{RGB}{255,217,179}
\definecolor{pastelred}{RGB}{255,179,186}
\definecolor{pastelcyan}{RGB}{186,225,255}
\definecolor{pastelyellow}{RGB}{255,255,186}
\definecolor{pastelblue}{RGB}{186,201,255}

\begin{tikzpicture}[scale=0.36]
    \newcommand{\bit}[2][]{    % Macro : dessine un bit et avance
        \draw[fill=#1] (cur) rectangle ++(1,1);
        \node at ($(cur)+(0.5,0.5)$) {#2};
        \coordinate (cur) at ($(cur)+(1,0)$);
    }

    \begin{scope}[yshift=0.5cm]
        \coordinate (cur) at (-1,0);
        \bit[white]{1}
        \bit[white]{0}
        \bit[white]{0}
        \bit[white]{0}
        \bit[white]{$\bar{1}$}
        \bit[white]{0}
        \bit[white]{$\bar{1}$}
        \bit[white]{0}
        \bit[white]{0}
        \bit[white]{0}
        \bit[white]{0}
        \bit[white]{$\bar{1}$}
        \bit[white]{0}
        \bit[white]{$\bar{1}$}
        \bit[white]{0}
        \bit[white]{0}
        \bit[white]{$\bar{1}$}
        \draw[dashed] (-4,-0.9) -- (16,-0.9);
    \end{scope}

    \begin{scope}[yshift=-2.4cm]
        \node[rotate=90, anchor=south] at (-5.6,0.5) {chunk};
        \node[rotate=90, anchor=south] at (-4.6,0.5) {divide};

        \coordinate (cur) at (-4,0);
        \bit[pastelyellow]{0}
        \bit[pastelyellow]{0}
        \bit[pastelyellow]{0}
        \bit[pastelyellow]{1}

        \coordinate (cur) at (2,0);
        \bit[pastelred]{0}
        \bit[pastelred]{$\bar{1}$}
        \bit[pastelred]{0}
        \bit[pastelred]{$\bar{1}$}

        \coordinate (cur) at (7,0);
        \bit[pastelyellow]{0}
        \bit[pastelyellow]{0}
        \bit[pastelyellow]{0}
        \bit[pastelyellow]{$\bar{1}$}

        \coordinate (cur) at (12,0);
        \bit[pastelcyan]{$\bar{1}$}
        \bit[pastelcyan]{0}
        \bit[pastelcyan]{0}
        \bit[pastelcyan]{$\bar{1}$}
    \end{scope}
\end{tikzpicture}
    \caption{Subdividing $60375_{10} = 1000\overline{1}0\overline{1}0000\overline{1}0\overline{1}00\overline{1}_{CSD}$ into CSD patterns of size $w=4$. Chunk divide leads to $M=\{\!\{1,\overline{1}0\overline{1},\overline{1},\overline{1}00\overline{1}\}\!\}$ and $S=\{1,101,1001\}$}
    \label{fig:subdivision_csd}
\end{figure}

From now on, patterns obtained through binary decompositions are called binary patterns, whereas patterns obtained through CSD decompositions are called CSD patterns.

The reconstruction we choose is the most naive we can think of: adding successively each CSD pattern shifted by its position in the target constant, and subtracting if its corresponding pattern in $M$ is negative.
In Section~\ref{sec:experiments}, we show that even with this simple reconstruction, our new CSD-based heuristic outperforms state of the art \cite{aksoy_multiplierless_2022}.

\section{3-step-optimal approach}
\label{sec:CSDe}

In previous approaches, one specific binary or CSD pattern set $S$ has been arbitrarily chosen among all suitable pattern sets.
This of course leads to sub-optimal results.
One could, for example, work around this limitation by combining the whole Synthesis + Divide + Reconstruct flow into one exact model using SAT.
However, this idea is not computationally feasible due to the size of the models.

An in-between solution is to enumerate all suitable binary or CSD pattern sets and run optimal Synthesis + Reconstruct solvers on each set. However, as there is an exponential number of pattern sets, we introduce a parameter $w$ to upper bound the number of bits in a pattern. Hence, optimality depends on $w$: 
Having enumerated and tried every suitable pattern set on at most $w$ bits does not guarantee that there is no pattern set with better quality on $w+1$ bits.
Compared to SAT, this procedure has the added benefit that it can stop after a timeout and yield the best solution found so far.

\subsection{Enumeration of pattern sets}

We first consider the problem of enumerating all binary or CSD pattern sets for one target constant, and explain how to extend it to the case of more than one constant at the end of the section. 
We reduce this problem to exact cover (XC), thus allowing us to use the DLX algorithm of Knuth \cite{taocp}.
An instance of XC is defined by a couple $(I,O)$ such that $I$ is a set of items and $O\subseteq {\cal P}(I)$ is a set of options.
XC aims at enumerating every option subset $E\subseteq O$ which is a partition of $I$, \textit{i.e.}, $\forall \ i\in I, \#\{o\in E : i\in o\} = 1$.
Given a target constant $b_{W-1}b_{W-2}\ldots b_0$ of $W$ bits (this can either be a binary or CSD representation) and a maximum pattern width $w$, we generate the XC instance $(I,O)$ as follows:
$I$ is the set of non-zero positions in the target constant, \textit{i.e.}, $I = \{i\in \llbracket 0,W \llbracket : b_i \neq 0\}$;
$O$ contains an option for each subset of $I$ such that the distance between the left most position and the right most position does not exceed $w$, \textit{i.e.}, $O=\{ S \subseteq I : \max S - \min S < w \}$.
This is illustrated with an example in Fig.~\ref{fig:XCCSD}.

\begin{figure}
    \centering
    $\begin{array}{lllllllll}
        & & b_6 & b_5 & b_4 & b_3 & b_2 & b_1 & b_0\\
        Options &&1 & 0 & 1 & 0 & \overline{1} & 0 & \overline{1}\\
        \hline
        a=\{6\} & \leadsto & 1 &  &  &  &  &  &  \\
        b=\{4\} & \leadsto &  &  & 1 &  &  &  &  \\
        c=\{2\} & \leadsto &  &  &  &  & \overline{1} &  &  \\
        d=\{0\} & \leadsto &  &  &  &  &  &  & \overline{1} \\
        e=\{6,4\} & \leadsto & 1 & 0 & 1 &  &  &  &  \\
        f=\{4,2\} & \leadsto &  &  & 1 & 0 & \overline{1} &  &  \\
        g=\{2,0\} & \leadsto &  &  &  &  & \overline{1} & 0 & \overline{1} \\
        h=\{6,2\} & \leadsto & 1 & 0 & 0 & 0 & \overline{1} &  &  \\
        i=\{4,0\} & \leadsto &  &  & 1 & 0 & 0 & 0 & \overline{1} \\
        j=\{6,4,2\} & \leadsto & 1 & 0 & 1 & 0 & \overline{1} &  &  \\
        k=\{4,2,0\} & \leadsto &  &  & 1 & 0 & \overline{1} & 0 & \overline{1} \\
    \end{array}$
    \caption{XC instance associated with target constant $1010\overline{1}0\overline{1}$ and $w=5$: $I = \{6,4,2,0\}$ and $O$ contains options $a$ to $k$.
    For each option, we display on the right its associated pattern.
    This instance has 10 solutions: $\{a,b,c,d\}$, $\{a,b,g\}$, $\{a,d,f\}$, $\{a,c,i\}$, $\{a,k\}$, $\{c,d,e\}$, $\{e,g\}$, $\{b,d,h\}$, $\{h,i\}$, and $\{d,j\}$.
    For each solution $X$, the multiset $M_X$ contains the patterns associated with its options and the set $S_X$ contains the unique absolute values of the elements of $M_X$.
    Two multisets can produce the same pattern set:
    here, for $\{a,b,g\}$, we have $M_{abg}=\{\!\{1, 1, \overline{1}0\overline{1}\}\!\}$ and $S_{abg}=\{1,101\}$, and for $\{c,d,e\}$, we have $M_{cde}=\{\!\{\overline{1}, \overline{1},101\}\!\}$ and $S_{cde} = \{1,101\} = S_{abg}$.
    %, {\em i.e.}, $\{1\}$, $\{1,101\}$, $\{1,10\overline{1}\}$, $\{1,1000\overline{1}\}$, $\{1,10\overline{1}0\overline{1}\}$, $\{1,101\}$, $\{101\}$, $\{1,1000\overline{1}\}$, $\{1000\overline{1}\}$, and $\{1,1010\overline{1}\}$, respectively.}
    }
    \label{fig:XCCSD}
\end{figure}

When there are $n > 1$ target constants $T_1,\ldots,T_n$, we build an instance $(I_j,O_j)$ for each $j\in \llbracket 1,n \rrbracket$. Then, we rename every item $i\in I_j$ to $i^j$ so that all item sets are disjoint.
Finally, we define the final XC instance $(\uplus_{j\in \llbracket 1,n \rrbracket} I_j,\uplus_{j\in \llbracket 1,n \rrbracket} O_j)$.

When $w = W$, there are $2^{\#I}$ options as every subset of $I$ is an option.
When $w < W$, we discard subsets that contain non-zeros at a distance larger than $w$, and the number of options is upper bounded by $W \cdot 2^{w}$ (when the target constant only contains 1s, there are $2^{w}$ options for each bit position, except for the last positions from $W-w+1$ to $W$).
Hence, the scalability of the problem of enumerating all pattern sets highly depends on both $w$ and the number of non-zero digits in the target constant.

\subsection{CSD patterns vs binary patterns}

There is no correspondence between binary patterns and CSD patterns.
Indeed, $\{11\}$ is a binary pattern set for the binary target $1111$ but its corresponding CSD pattern set $\{10\overline{1}\}$ does not cover the corresponding CSD target $1000\overline{1}$.
Reciprocally, $\{10\overline{1}\}$ is a CSD pattern set for the CSD target $10\overline{1}0\overline{1}01$ but its corresponding binary pattern set $\{11\}$ does not cover the corresponding binary target $101101$.

Because the number of enumerated pattern sets highly depends on the number of non-zero digits of the target constants, we can experimentally verify that CSD pattern sets are considerably less numerous than binary pattern sets.
Fig.~\ref{fig:csd_vs_bin_split} verifies it, by showing the distribution of the average quantity of patterns sets of size four and five over all target constants of size 20 sorted by Hamming weight.
This illustrates how high the number of binary pattern sets depends on the Hamming weight.
Moreover, with large Hamming weights, the number of non-zero digits in CSD form is low, so there are fewer corresponding CSD pattern sets.

\begin{figure}
    \centering
    \begin{tikzpicture}[scale=0.8]
        \begin{axis}[
            width=11cm,
            height=6.2cm,
            xlabel={Hamming weight},
            ylabel={Number of pattern sets},
            legend style={
                at={(0.02,1)},
                anchor=north west,
                fill=white,
            },
            xmin=2, xmax=20,
            ymin=1, ymax=20000,
            ymode=log,
            grid=major,
        ]
            
            \addplot[orange!85!black, dotted, line width=1.2pt, mark=triangle*, mark size=2.2pt] table [x index=0, y index=1, col sep=space] {data/hamming_exhaustive_bin_w5.dat};
            \addlegendentry{BIN (w=5)}
            \addplot[red!80!black, dotted, line width=1.2pt, mark=square*, mark size=2.2pt] table [x index=0, y index=1, col sep=space] {data/hamming_exhaustive_bin_w4.dat};
            \addlegendentry{BIN (w=4)}
            
            \addplot[blue!75!black, solid, line width=1.4pt, mark=*, mark size=2.0pt] table [x index=0, y index=1, col sep=space] {data/hamming_exhaustive_csd_w5.dat};
            \addlegendentry{CSD (w=5)}
            \addplot[green!65!black, solid, line width=1.4pt, mark=diamond*, mark size=2.2pt] table [x index=0, y index=1, col sep=space] {data/hamming_exhaustive_csd_w4.dat};
            \addlegendentry{CSD (w=4)}
        \end{axis}
    \end{tikzpicture}
    \caption{Mean number of different pattern sets for every 20-bit constant %of size 20 bits 
    w.r.t Hamming weight.
    With increasing pattern bit width (from 4 to 5 bits), the number of patterns to explore grows exponentially for binary encoding but stays reasonable for CSD.}
    \label{fig:csd_vs_bin_split}
\end{figure}

\iffalse
\begin{figure}
    \centering
    \begin{tikzpicture}[scale=0.8]
        \begin{axis}[
            width=11cm,
            height=7cm,
            xlabel={Value of the target constant},
            ylabel={Number of pattern sets},
            legend pos=north west,
            xmin=4097, xmax=12137,
            ymin=0, ymax=160,
            grid=major,
        ]
        \addplot[ybar, bar width=0.1pt, draw=red] table [x index=0, y index=1, col sep=semicolon] {data/exhaustive_comparison_tronque.dat};
        \addlegendentry{BIN}
        \addplot[ybar, bar width=0.5pt, draw=orange] table [x index=0, y index=2, col sep=semicolon] {data/exhaustive_comparison_tronque.dat};
        \addlegendentry{CSD}
        \end{axis}
    \end{tikzpicture}
    \caption{Distribution of the number of pattern sets}
    \label{fig:csv_vs_bin2}
\end{figure}
\fi

\section{Experimental evaluation}
\label{sec:experiments}

The proposed models and approaches are available as an open-source tool-flow \url{https://gitlab.inria.fr/emeraude/vlcm}.
We first analyze the global quality of the proposed methods and obtained solutions with respect to Aksoy et al. \cite{aksoy_multiplierless_2022} which is the state-of-the-art reference. Then, we use our 3-step optimal approach to assess the optimization potential of different pattern configurations. Finally, we motivate the further work towards better optimization metrics based on FPGA synthesis results.

To run our experiments, we used 3 benchmark sets:
\begin{itemize}
    \item Cryptographic: consists of 14 single target constant sets with word length from 204 bits to 751 bits, extracted from \cite{Bernstein2014} and \cite{Defeo2018};
    \item DSP: consists of 120 instances (3 to 9 coefficients per target set) of 53 bits. These are coefficients of ANSI S1.11-compatible 1/3 octave digital filters designed in FP64 and typically unstable to quantize to smaller fixed-point formats;
    \item Random: consists of 25 instances for every word length in $\{32, 64, 128, 256\}$ with 10 randomly generated target constants by set (100 instances in total).
\end{itemize}

Solving SAT instances was limited to max. 600\,s per instance using the CaDiCaL solver by Biere et al.~\cite{biere_cadical_2024}.
If the SAT solver failed to improve upon the heuristic decomposition/reconstruction, the heuristic solution is used instead.

\subsection{Quality of the proposed SAT-based and heuristic solutions}

First, we analyze whether the SAT-based pattern decomposition and reconstruction can be used for practical VLCM instances.
Figure~\ref{fig:sat_vs_aksoy_all_mcm} shows a comparison of the total adder counts across all benchmarks.
Missing results are due to timeouts of the SAT solver.
It can be seen that SAT usually reduces costs if it finds a solution within the set timeout. The relative improvement goes up to 57\% for some instances, and the absolute adder savings of up to 68 adders have been reported.
These savings are mainly due to an improved reconstruction as illustrated in Fig.~\ref{fig:reconstruct_comparison}, where we ran the SAT reconstruction on Aksoy's patterns. We note that half of the TOLL-deduced patterns was not needed for reconstruction. This can happen in the proposed SAT approach too, though to a much smaller extent, due to the inherent sub-optimality of the 3-step approach. 

\begin{figure}[tbp]
    \centering
    \includegraphics[height=6cm]{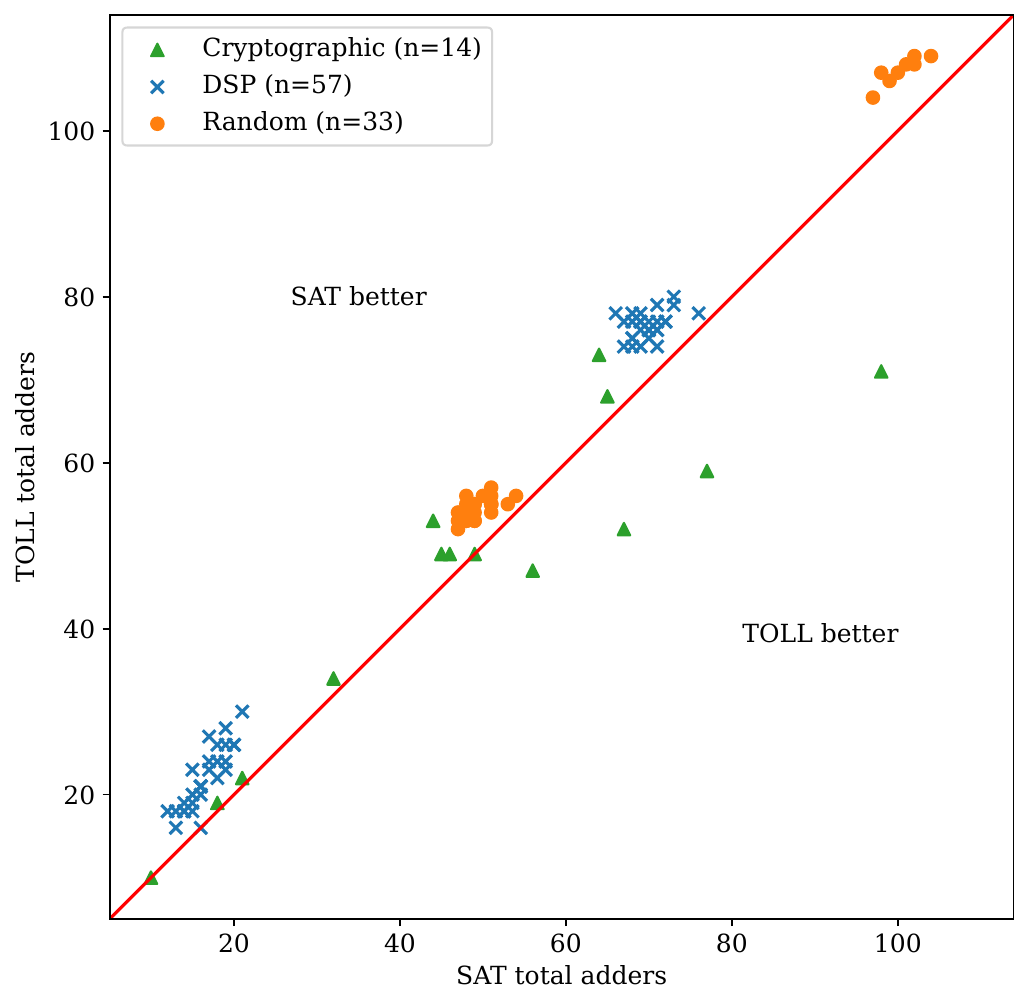}
    \caption{SAT (MCM) vs. TOLL~\cite{aksoy_multiplierless_2022} for pattern size 12 bits across all benchmarks.}
    \label{fig:sat_vs_aksoy_all_mcm}
\end{figure}

\begin{figure}[htbp]
    \centering
    \begin{subfigure}[b]{0.5\textwidth}
        \centering
        \scalebox{0.7}{%
    \begin{tikzpicture}[scale=0.2]
        \addnode{-35,-8}{add0}{0}{9}{0}
        \addnode{-46,-8}{add1}{0}{13}{0}
        \addnode{-27,-8}{add2}{15}{0}{0}
        \addnode{-21,-17}{add3}{0}{29}{0}
        \addnode{-6,-8}{add4}{12}{0}{0}
        \addnode{0,-17}{add5}{0}{24}{0}
        \addnode{-13,-26}{add6}{0}{37}{0}
        \addnode{-41,-17}{add7}{27}{0}{0}
        \addnode{6,-26}{add8}{0}{37}{0}
        \draw (-48,0) node(x351) {$351$};
        \draw (-44,0) node(x1) {$1$};
        \draw (-37,0) node(x901) {$901$};
        \draw (-31,1) node(x467) {$467$};
        \draw[hwbus] (x467.south) -- ++(0,-1) coordinate (x467);
        \draw (-25,0) node(x3461) {$3461$};
        \draw (-19,0) node(x95) {$95$};
        \draw (-11,0) node(x3) {$3$};
        \draw (-8,0) node(x1509) {$1509$};
        \draw (-4,0) node(x1025) {$1025$};
        \draw (2,0) node(x465) {$465$};
        \draw (8,0) node(x5) {$5$};
        
        \draw (add0o) ++(0,-2) node {$3826565$};
        \draw (add1o) ++(0,-2) node {$863$};
        \draw (add2o) ++(0,-2) node {$15306117$};
        \draw (add3o) ++(0,-2) node {$51018042757$};
        \draw (add4o) ++(0,-2) node {$6181889$};
        \draw (add5o) ++(-1,-2) node {$7807587329$};
        \draw[hwbus, ->] (add6o) -- ++(0,-2) node[below] {$\textbf{463334903173}$};
        \draw[hwbus, ->] (add7o) -- ++(0,-2) node[below] {$\textbf{115833725829}$};
        \draw[hwbus, ->] (add8o) -- ++(0,-2) node[below] {$\textbf{695002354689}$};
    
        \draw[hwbus] (x901.south) -| (add0i0);
        \draw[hwbus] (x467.south) -| (add0i1);
        \draw[hwbus] (x351.south) -| (add1i0);
        \draw[hwbus] (x1.south) -| (add1i1);
        \draw[hwbus] (x467.south) -| (add2i0);
        \draw[hwbus] (x3461.south) -| (add2i1);
        \draw[hwbus] (add2o) -| (add3i0);
        \draw[hwbus] (x95.south) -| (add3i1);
        \draw[hwbus] (x1509.south) -| (add4i0);
        \draw[hwbus] (x1025.south) -| (add4i1);
        \draw[hwbus] (add4o) -| (add5i0);
        \draw[hwbus] (x465.south) -| (add5i1);
        \draw[hwbus] (add3o) -| (add6i0);
        \draw[hwbus] (x3.south) -| (add6i1);
        \draw[hwbus] (add1o) -| (add7i0);
        \draw[hwbus] (add0o) -| (add7i1);
        \draw[hwbus] (add5o) -| (add8i0);
        \draw[hwbus] (x5.south) -| (add8i1);
    \end{tikzpicture}
}
        \caption{Unique patterns and reconstruction by TOLL~\cite{aksoy_multiplierless_2022} }
        \label{fig:aksoy}
    \end{subfigure}
    \vspace{-0.6em}
    \begin{subfigure}[b]{0.5\textwidth}
        \centering
        \scalebox{0.7}{%
    \begin{tikzpicture}[scale=0.2]
        \addnode{-5,-8}{add0}{8}{0}{0}
        \addnode{0,-17}{add1}{14}{0}{0}
        \addnode{-10,-26}{add2}{15}{0}{0}
        \addnode{8,-26}{add3}{0}{13}{0}
        \subnode{-1,-32}{sub4}{0}{0}{-1}
        \addnode{14,-45}{add5}{2}{0}{0}
        \draw (-12,0) node(x3461) {$3461$};
        \draw (-7,0) node(x3) {$3$};
        \draw (-3,0) node(x95) {$95$};
        \draw (2,0) node(x467) {$467$};
        \draw (10,0) node(x901) {$901$};
        \draw (16,0) node(x1) {$1$};

        \draw (-22,0) node(x351) {$351$};
        \draw (-17,0) node(x1509) {$1509$};
        \draw (21,0) node(x1025) {$1025$};
        \draw (26,0) node(x465) {$465$};
        \draw (31,0) node(x5) {$5$};
        
        \draw (add0o) ++(-1,-2) node {$863$};
        \draw (add1o) ++(-1,-2) node {$14139859$};
        \draw (sub4o) ++(0,-2) node {$173750588672$};
        \draw[hwbus, ->] (add3o) -- ++(0,-2) node[below] {$\textbf{115833725829}$};
        \draw[hwbus, ->] (add2o) -- ++(0,-2) node[below] {$\textbf{463334903173}$};
        \draw[hwbus, ->] (add5o) -- ++(0,-2) node[below] {$\textbf{695002354689}$};
        
        \draw[hwbus] (x3.south) -| (add0i0);
        \draw[hwbus] (x95.south) -| (add0i1);
        \draw[hwbus] (add0o) -| (add1i0);
        \draw[hwbus] (x467.south) -| (add1i1);
        \draw[hwbus] (x3461.south) -| (add2i0);
        \draw[hwbus] (add1o) -| (add2i1);
        \draw[hwbus] (add1o) -| (add3i0);
        \draw[hwbus] (x901.south) -| (add3i1);
        \draw[hwbus] (add2o) -| (sub4i0);
        \draw[hwbus] (add3o) -| (sub4i1);
        \draw[hwbus] (sub4o) -| (add5i0);
        \draw[hwbus] (x1.south) -| (add5i1);
    \end{tikzpicture}
}
        \caption{Optimal reconstruction on TOLL-issued patterns}
        \label{fig:nicolai}
    \end{subfigure}
    \caption{Comparison of TOLL~\cite{aksoy_multiplierless_2022} reconstruction (9 adders) with the proposed MCM technique (6 adders and 5 patterns remain unused) on \texttt{butterworth\_3\_3\_b} instance.}
    \label{fig:reconstruct_comparison}
\end{figure}

The proposed CSD heuristic was able to solve all instances quickly, as shown in Table~\ref{tab:rel-impr-aksoy}.
Due to time considerations, SAT was launched for $w=4$ and $w=8$ only for DSP.
It can be seen that using CSD for VLCM leads to significant savings, especially for larger pattern sizes.

\begin{table}[t]
    \centering
    \caption{Relative improvement statistics w.r.t. TOLL~\cite{aksoy_multiplierless_2022}. SAT statistics are computed only over instances that finished.}
    \label{tab:rel-impr-aksoy}
    \begin{tabular}{ll rr r}
        \toprule
        benchmark & w & \multicolumn{1}{c}{CSD} & \multicolumn{2}{c}{SAT} \\
        \cmidrule(lr){3-3} \cmidrule(lr){4-5}
        & & mean (\%) & solved & mean (\%) \\
        \midrule
            \multirow{3}{*}{DSP}
            & 4 & 7.6 & 30/120 & 15.6 \\
            & 8 & 14.4 & 21/120 & 28.5 \\
            & 12 & 15.6 & 57/120 & 16.6 \\
        \midrule
        \multirow{3}{*}{Cryptographic}
            & 4 & 5.2 & -- & -- \\
            & 8 & 13.4 & -- & -- \\
            & 12 & 16.0 & 14/14 & -3.8 \\
        \midrule
        \multirow{3}{*}{Random}
            & 4 & -10.3 & -- & -- \\
            & 8 & 11.1 & -- & -- \\
            & 12 & 10.8 & 33/100 & 8.7 \\
        \bottomrule
    \end{tabular}
\end{table}

\begin{figure}[tbp]
    \centering
    \includegraphics[width=.9\linewidth]{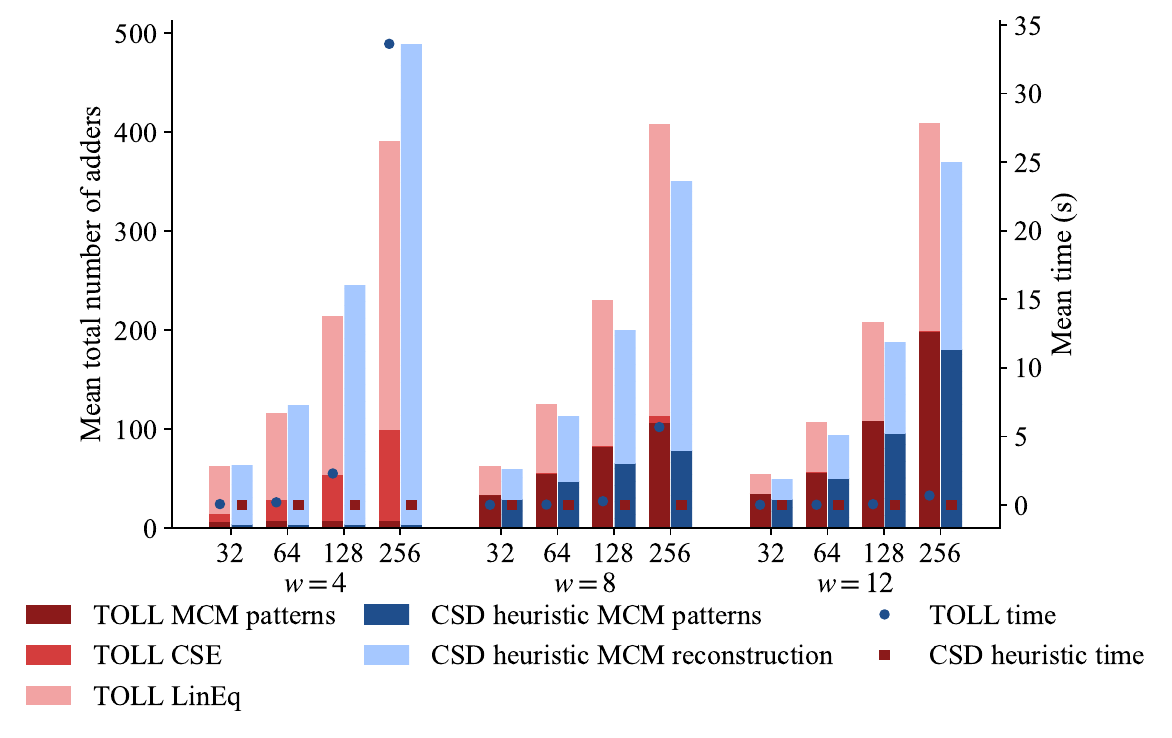}
    \caption{Mean number of adders and timings, per word length, for the Steps 1 and 2 on the Random benchmark. TOLL reconstruction includes linear equations (LinEq) and CSE. }
    \label{fig:aksoy_vs_naive}
\end{figure}

Figure~\ref{fig:aksoy_vs_naive} indicates one of the sources of these savings is that CSE fails beyond $w=4$.
However, savings for CSD-based heuristic also come from improved pattern decomposition and hence, lower cost of MCM over patterns.

\subsection{Assessing optimization potential for different strategies}

Section~\ref{sec:CSDe} proposed a 3-step optimal solution based on enumeration of all pattern sets, whether binary or CSD. The exponential number of pattern sets and the exponential complexity of the reconstruction currently make it infeasible for practical instances. Nevertheless, exploring the complete design space and assessing the 3-step approaches on small-scale examples remains informative, in particular to estimate the distance between the proposed and reference approaches to the actual minima.

To this end, we created a small random benchmark with 10 target constants for each word length from 12 to 25 bits. The resolution time for optimal reconstruction is essentially exponential: 15-bit targets can be solved in a few seconds, 19-bit in thousands of seconds, and size 25 bits in millions.

We observe that the search space for patterns is not only large but unevenly distributed, spanning multiple possibilities for the total number of adders (see Fig.~\ref{fig:histogram_num_adders}). Interestingly, here the CSD heuristic got optimal solution. The non-optimality of the SAT solution highlights the effect of disjoint treatment of each step of the VLCM method as well as the superiority of CSD over binary number representations.

To assess the optimization potential of the TOLL tool and the proposed approaches, we enumerate the design space using the CSD-based patterns and, for fairness, retain the best result for different pattern word lengths for each strategy. Figure~\ref{fig:enumeration_vs_aksoy} shows that the proposed approaches outperform the reference one, while the CSD heuristic is often close to the optimum.
Its advantage over SAT can again be explained by the difference in number formats for partitioning.

Overall, this trend supports the hypothesis that optimizing the total number of patterns is a good proxy metric.

\begin{figure}[t!]
    \centering
    \begin{tikzpicture}[scale=1]
        \begin{axis}[
            width=8cm,
            height=3.5cm,
            xlabel={Total number of adders},
            ylabel={\shortstack{\#CSD patterns}},
            xmin=5, xmax=12,
            xtick={6,7,8,9},
            ymin=0, ymax=1000,
            grid=major,
            legend style={at={(1,.9)},align=left, font =\scriptsize}
        ]
            \addplot[hist={bins=4,data min=5.5,data max=9.5}, fill=blue!20, draw=blue!40, forget plot] table [y index=0] {data/48863421_w7_hist_quality.dat};
            \addplot[red, very thick] coordinates {(8,0) (8,1000)};
            \addlegendentry{TOLL~\cite{aksoy_multiplierless_2022}}
            \addplot[ForestGreen, very thick] coordinates {(7,0) (7,1000)};
            \addlegendentry{SAT}
            \addplot[blue, very thick] coordinates {(6,0) (6,1000)};
            \addlegendentry{CSD heuristic}
        \end{axis}
    \end{tikzpicture}
    \caption{Design space exploration for multiplication by 48863421 (26 bits) and word length of pattern set to 5. Vertical bars indicate the solution returned by each strategy.}
    \label{fig:histogram_num_adders}
\end{figure}

\begin{figure}[btp]
    \centering
    \begin{tikzpicture}[scale=0.7]
        \begin{axis}[
            width=12cm,
            height=7cm,
            xlabel={Target constant size},
            ylabel={\shortstack{Average of the total\\number of adders}},
            ymin=3, ymax=10,
            xmin=11, xmax=26,
            xtick={12,13,...,25},
            ymajorgrids,
            legend style={
                at={(0.02,1.4)},
                anchor=north west,
                font=\scriptsize
            },
            unbounded coords=discard
        ]
            \addplot[name path=lower, draw=none, forget plot] table[x index=0, y index=1] {data/compare_quality.dat};
            \addplot[name path=upper, draw=none, forget plot] table[x index=0, y index=2] {data/compare_quality.dat};
            \addplot[blue!20] fill between[of=lower and upper];
            \addlegendentry{CSD enumeration}
            
            \addplot[red,only marks,mark=*] table [x expr=\thisrowno{0}+0.15, y index=3] {data/compare_quality.dat};
            \addlegendentry{TOLL~\cite{aksoy_multiplierless_2022} (Best of 4-8-12)}

            \addplot[ForestGreen,only marks,mark=triangle*] table [x expr=\thisrowno{0}, y index=5] {data/compare_quality.dat};
            \addlegendentry{SAT (Best of 4-8-12)}
            
            \addplot[blue,only marks,mark=square*] table [x expr=\thisrowno{0}-0.15, y index=4] {data/compare_quality.dat};
            \addlegendentry{CSD heuristic (Best of 4-8-12)}
        \end{axis}
    \end{tikzpicture}
    \caption{Mean values of the total number of adders returned by different strategies over the random benchmark. The light blue shape is the exhaustively enumerated design-space.}
    \label{fig:enumeration_vs_aksoy}
\end{figure}

\subsection{Some motivating synthesis results}

Table~\ref{tab:synth_butter} shows synthesis results for \texttt{butter\-worth\_3\_3\_b} from the DSP benchmark for 49 bit coefficient word size and 16 bit input word size using Vivado v2024.2 and pattern size $w \in \{4, 12\}$ targeting a Virtex Ultrascale+ (\texttt{xcvu9p}\texttt{-}\texttt{flga2104}\texttt{-}\texttt{2L}\texttt{-}\texttt{e}).
The MCM circuits are sandwiched between registers for clock speed optimization; the target frequency is varied with $0.125\,\text{ns}$ steps and we report results for the designs with maximum clock speed.

To no surprise, the standard multipliers with enabled DSP blocks have the smallest Look-Up Table (LUT) utilization and highest clock speeds.
CSD and TOLL yield LUT costs between SAT and Vivado.
The results further show that the adder count reduction for $w=12$ compared to $w=4$ also translates to resource savings for FPGA implementation.

Interestingly, adder costs do not directly correlate with LUT costs.
A closer correlation is given when counting the number of one-bit adders as given in column \enquote{Bit-add}.
For example, TOLL and SAT with $w=12$ have equal LUT costs despite an adder count difference of four; and CSD has less adders than TOLL and SAT for $w=4$, but higher LUT costs.
The reason might lie in very different width of adders in the top of the circuit when computing patterns, and those used at reconstruction.
This motivates further research to also include bit-level costs into the VLCM optimization.

\begin{table}[tbp]
    \centering
    \caption{Synthesis results for \texttt{butterworth\_3\_3\_b}; Vivado\textsuperscript{$\dagger$}: with enabled embedded multipliers}
    \label{tab:synth_butter}
    \begin{tabular}{lccccc}
        \toprule
        & $w$ & \#\,Add & Bit-add & \#\,LUTs $\downarrow$ & $f_\text{max}$, MHz $\uparrow$ \\
        \midrule
        Vivado\textsuperscript{$\dagger$} & -- & -- & -- & 51 + 6 DSPs & 444.44 \\
        Vivado & -- & -- & -- & 945 & 258.06 \\
        \midrule
        TOLL \cite{aksoy_multiplierless_2022} & 4 & 30 & 580 & 544 & 210.53 \\
        CSD & 4 & \textbf{24} & 644 & 634 & \textbf{228.57} \\
        SAT & 4 & 26 & \textbf{471} & \textbf{468} & 170.21 \\
        \midrule
        TOLL \cite{aksoy_multiplierless_2022} & 12 & 20 & \textbf{448} & \textbf{416} & 195.12 \\
        CSD & 12 & 21 & 539 & 532 & \textbf{242.42} \\
        SAT & 12 & \textbf{16} & 474 & \textbf{416} & 222.22 \\
        \bottomrule
    \end{tabular}
\end{table}

\section{Conclusion and Perspectives}

This paper introduced exact and heuristic methods solving the VLCM problem and proposed an exhaustive design-space exploration to assess the relevance of chosen optimization metrics. 
Our approaches outperform state of the art algorithms in terms of adder cost due to a better choice of patterns, which yield smaller MCMs on Step 2, and thanks to the optimal reconstruction on Step 3, but optimal SAT approach has prohibitive run times. Considering CSD representation proved to be a good choice to provide a good quality of a fast heuristic. 

As perspective, extending the SAT models to operate on CSD instead of binary decomposition might be a good choice, though with an unknown impact on the run time. While the model gets more complex ($\uparrow$ run time), the solver might converge with decreased iterations ($\downarrow$ run time).
Another improvement for the reconstruction could be through restricting carries in the adder blocks, which will simplify the model only slightly impacting the quality. 
Furthermore, doing a fully-heuristic dedicated CSD-based reconstruction could provide a good scalability/quality trade-off. Overall, statistical analysis of small-scale pattern enumeration results may help characterize patterns that are likely to produce efficient shift-and-add implementations.

Targeting finer-grained hardware-aware metrics, e.g. one-bit adder costs, or using compressor trees~\cite[Chapter~7]{de_dinechin_application-specific_2024} could be beneficial for both optimal and heuristic reconstruction parts. 
An integration into FloPoCo~\cite{de_dinechin_designing_2011} would enable the use of our proposed algorithms into existing workflows while also giving direct access to optimized compressor tree implementations.

\bibliographystyle{IEEEtran}
\bibliography{bibliography.bib}

@inproceedings{garcia_mcm_dnn_2025,
    author = {R{\'{e}}mi Garcia and L{\'{e}}o Pradels and Silviu{-}Ioan Filip and Olivier Sentieys},
    title = {Hardware-Aware Training for Multiplierless Convolutional Neural Networks},
    booktitle = {32nd Symposium on Computer Arithmetic, ARITH},
    pages = {9--16},
    publisher = {IEEE},
    year = {2025}
}

@book{de_dinechin_application-specific_2024,
	address = {Cham},
	title = {Application-Specific Arithmetic: Computing Just Right for the Reconfigurable Computer and the Dark Silicon Era},
	publisher = {Springer International Publishing},
	author = {De Dinechin, Florent and Kumm, Martin},
	year = {2024}
}

@article{de_dinechin_designing_2011,
	title = {Designing Custom Arithmetic Data Paths with FloPoCo},
	volume = {28},
	number = {4},
	journal = {IEEE Design \& Test of Computers},
	author = {de Dinechin, Florent and Pasca, Bogdan},
	year = {2011},
	pages = {18--27}
}

@article{GarridoM21,
    author = {Mario Garrido and Pedro Malag{\'{o}}n},
    title = {The Constant Multiplier FFT},
    journal = {IEEE Trans. Circuits Syst. I Regul. Pap.},
    volume = {68},
    number = {1},
    pages = {322--335},
    year = {2021}
}

@inproceedings{AksoyFM14,
  author = {Levent Aksoy and Paulo F. Flores and Jos{\'{e}} Monteiro},
  title = {Efficient design of FIR filters using hybrid multiple constant multiplicationson FPGA},
  booktitle = {32nd International Conference on Computer Design},
  pages = {42--47},
  publisher = {IEEE},
  year = {2014}
}

@inproceedings{Volkova0DK23,
    author = {Anastasia Volkova and R{\'{e}}mi Garcia and Florent de Dinechin and Martin Kumm},
    title = {Hardware-Optimal Digital FIR Filters: One ILP to Rule Them all and in Faithfulness Bind Them},
    booktitle = {57th Asilomar Conference on Signals, Systems, and Computers},
    pages = {1574--1578},
    publisher = {IEEE},
    year = {2023}
}

@inproceedings{HabermannKKV22,
    author = {Tobias Habermann and Jonas K{\"{u}}hle and Martin Kumm and Anastasia Volkova},
    title = {Hardware-Aware Quantization for Multiplierless Neural Network Controllers},
    booktitle = {Asia Pacific Conference on Circuit and Systems},
    pages = {541--545},
    publisher = {IEEE},
    year = {2022}
}

@article{mcm_iir_2022,
    author = {R{\'{e}}mi Garcia and Anastasia Volkova and Martin Kumm and Alexandre Goldsztejn and Jonas K{\"{u}}hle},
    title = {Hardware-Aware Design of Multiplierless Second-Order IIR Filters With Minimum Adders},
    journal = {IEEE Trans. Signal Process.},
    volume = {70},
    pages = {1673--1686},
    year = {2022}
}

@inproceedings{HardieckHWMKZ23,
    author = {Martin Hardieck and Tobias Habermann and Fabian Wagner and Michael Mecik and Martin Kumm and Peter Zipf},
    title = {More AddNet: A deeper insight into DNNs using FPGA-optimized multipliers},
    booktitle = {International Symposium on Circuits and Systems},
    pages = {1--5},
    publisher = {IEEE},
    year = {2023}
}

@article{dempster_use_1995,
	title = {Use of minimum-adder multiplier blocks in {FIR} digital filters},
	volume = {42},
	number = {9},
	journal = {IEEE Transactions on Circuits and Systems II: Analog and Digital Signal Processing},
	author = {Dempster, A.G. and Macleod, M.D.},
	year = {1995},
	pages = {569--577}
}

@article{kumm_optimal_2018,
	title = {Optimal Constant Multiplication Using Integer Linear Programming},
	volume = {65},
	number = {5},
	journal = {IEEE Transactions on Circuits and Systems II: Express Briefs},
	author = {Kumm, Martin},
	year = {2018},
	pages = {567--571},
}

@article{hartley_subexpression_1996,
	title = {Subexpression sharing in filters using canonic signed digit multipliers},
	volume = {43},
	number = {10},
	journal = {Transactions on Circuits and Systems II: Analog and Digital Signal Processing},
	publisher = {IEEE},
	author = {Hartley, R.I.},
	year = {1996},
	pages = {677--688}
}

@article{voronenko_multiplierless_2007,
	title = {Multiplierless multiple constant multiplication},
	volume = {3},
	number = {2},
	journal = {ACM Transactions on Algorithms},
	author = {Voronenko, Yevgen and Püschel, Markus},
	year = {2007}
}

@inproceedings{aksoy_optimization_2007, 
    year = {2007}, 
    author = {Aksoy, Levent and Costa, Eduardo and Flores, Paulo and Monteiro, Jose}, 
    title = {Optimization of area in digital FIR filters using gate-level metrics}, 
    booktitle = {44th ACM/IEEE Design Automation Conference},
    pages = {420--423},
}

@article{lou_fine-grained_2015, 
    year = {2015}, 
    title = {Fine-Grained Critical Path Analysis and Optimization for Area-Time Efficient Realization of Multiple Constant Multiplications}, 
    author = {Lou, Xin and Yu, Ya Jun and Meher, Pramod Kumar}, 
    journal = {IEEE Transactions on Circuits and Systems I: Regular Papers},
    pages = {863 -- 872}, 
    number = {3}, 
    volume = {62}, 
}

@INPROCEEDINGS{johansson_bit-level_2007,
  author = {Johansson, K. and Gustafsson, O. and Wanhammar, L.},
  booktitle = {2007 14th IEEE International Conference on Electronics, Circuits and Systems}, 
  title = {Bit-Level Optimization of Shift-and-Add Based FIR Filters}, 
  year = {2007},
  volume = {},
  number = {},
  pages = {713-716}
}

@INPROCEEDINGS{johansson_detailed_2005,
  author = {Johansson, K. and Gustafsson, O. and Wanhammar, L.},
  booktitle = {Proceedings of the 2005 European Conference on Circuit Theory and Design}, 
  title = {A detailed complexity model for multiple constant multiplication and an algorithm to minimize the complexity}, 
  year = {2005},
  volume = {3},
  number = {},
  pages = {465--468}
}

@article{aksoy_search_2010,
	title = {Search algorithms for the multiple constant multiplications problem: Exact and approximate},
	volume = {34},
	number = {5},
	journal = {Microprocessors and Microsystems},
	author = {Aksoy, Levent and Güneş, Ece Olcay and Flores, Paulo},
	year = {2010},
	pages = {151--162}
}

@inproceedings{biere_cadical_2024,
	title = {CaDiCaL 2.0},
	booktitle = {Computer Aided Verification},
	publisher = {Springer Nature Switzerland},
	author = {Biere, Armin and Faller, Tobias and Fazekas, Katalin and Fleury, Mathias and Froleyks, Nils and Pollitt, Florian},
	editor = {Gurfinkel, Arie and Ganesh, Vijay},
	year = {2024},
	pages = {133--152}
}

@book{taocp,
    author = {Donald Ervin Knuth},
    title = {The art of computer programming, Volume 4B,  Combinatorial Algorithms, Part 2},
    publisher = {Addison-Wesley},
    year = {2022},
}

@ARTICLE{garcia23,
    author = {Garcia, Rémi and Volkova, Anastasia},
    journal = {IEEE Transactions on Circuits and Systems I: Regular Papers}, 
    title = {Toward the Multiple Constant Multiplication at Minimal Hardware Cost}, 
    year = {2023},
    volume = {70},
    number = {5},
    pages = {1976-1988}
}

@ARTICLE{fiege24,
    author = {Fiege, Nicolai and Kumm, Martin and Zipf, Peter},
    journal = {IEEE Transactions on Circuits and Systems I: Regular Papers}, 
    title = {Bit-Level Optimized Constant Multiplication Using Boolean Satisfiability}, 
    year = {2024},
    volume = {71},
    number = {1},
    pages = {249-261}
}

@inproceedings{cantaloube25,
    TITLE = {A new Constraint Programming model for the Multiple Constant Multiplication},
    AUTHOR = {Cantaloube, Th{\'e}o and Peng, Xiao and Solnon, Christine and Volkova, Anastasia},
    BOOKTITLE = {ModRef - 24th workshop on Constraint Modelling and Reformulation},
    YEAR = {2025}
}

@INPROCEEDINGS{kumm12,
    author = {Kumm, Martin and Zipf, Peter and Faust, Mathias and Chang, Chip-Hong},
    booktitle = {IEEE International Symposium on Circuits and Systems}, 
    title = {Pipelined adder graph optimization for high speed multiple constant multiplication}, 
    year = {2012},
    volume = {},
    number = {},
    pages = {49-52}
}

@article{aksoy_multiplierless_2022,
    title = {Multiplierless Design of Very Large Constant Multiplications in Cryptography},
    volume = {69},
    number = {11},
    journal = {Transactions on Circuits and Systems II: Express Briefs},
    publisher = {IEEE},
    author = {Aksoy, Levent and Roy, Debapriya Basu and Imran, Malik and Karl, Patrick and Pagliarini, Samuel},
    year = {2022},
    pages = {4503--4507}
}

@inproceedings{aksoy_multiplierless_2024,
	title = {Multiplierless Design of High-Speed Very Large Constant Multiplications},
	booktitle = {29th Asia and South Pacific Design Automation Conference},
	author = {Aksoy, Levent and Roy, Debapriya Basu and Imran, Malik and Pagliarini, Samuel},
	year = {2024},
	pages = {957--962}
}

@misc{Bernstein2014,
    author = {Bernstein, Daniel J. and Lange, Tanja},
    title = {SafeCurves: Choosing Safe Curves for ECC},
    year = {2014}
}

@misc{Defeo2018,
    author = {De Feo, Luca and Jao, David and Plût, Jérôme},
    title = {Supersingular Isogeny Key Encapsulation},
    year = {2018}
}

@article{Cappello1984,
    author = {Peter Cappello and Kenneth Steiglitz},
    title = {Some Complexity Issues in Digital Signal Processing},
    journal = {IEEE Transactions on Acoustics, Speech, and Signal Processing},
    volume = {32},
    number = {5},
    pages = {1037--1041},
    year = {1984}
}

@article{Rafferty2017,
    author = {C. Rafferty and M. O'Neill and N. Hanley},
    title = {Evaluation of Large Integer Multiplication Methods on Hardware},
    journal = {Transactions on Computers},
    volume = {66},
    number = {8},
    pages = {1369--1382},
    year = {2017},
    publisher = {IEEE}
}

@article{Chaves2007,
    author = {R. Chaves and L. Sousa},
    title = {Improving Residue Number System Multiplication with More Balanced Moduli Sets and Enhanced Modular Arithmetic Structures},
    journal = {Computers \& Digital Techniques},
    volume = {1},
    number = {5},
    pages = {472--480},
    year = {2007},
    publisher = {IET}
}

@inproceedings{Hartley1991,
    author = {R. Hartley},
    title = {Optimization of Canonic Signed Digit Multipliers for Filter Design},
    booktitle = {Proceedings of the International Symposium on Circuits and Systems},
    year = {1991},
    pages = {1992--1995},
    volume = {4},
    publisher = {IEEE}
}

@article{Dempster1994,
    author = {Andrew G. Dempster and Malcolm D. Macleod},
    title = {Constant Integer Multiplication Using Minimum Adders},
    journal = {IEE Proceedings - Circuits, Devices and Systems},
    volume = {141},
    number = {5},
    pages = {407--413},
    year = {1994},
    publisher = {IET}
}

@incollection{Reitwiesner1960,
    author = {George W. Reitwiesner},
    title = {Binary Arithmetic},
    booktitle = {Advances in Computers},
    editor = {Franz L. Alt},
    volume = {1},
    pages = {231--308},
    year = {1960},
    publisher = {Academic Press}
}

@article{Thong2009,
    author = {J. Thong and N. Nicolici},
    title = {Time-Efficient Single Constant Multiplication Based on Overlapping Digit Patterns},
    journal = {IEEE Transactions on Very Large Scale Integration Systems},
    year = {2009},
    volume = {17},
    number = {9},
    pages = {1353--1357},
}

\end{document}